\pdfoutput=1
\documentclass[journal]{IEEEtran}
\usepackage[thinlines]{easytable}
\usepackage{mathtools,amssymb,bm}
\usepackage{amsmath}
\usepackage{cite}
\usepackage{graphicx}
\usepackage{epstopdf}

\usepackage{cite}
\newtheorem{assumption}{Assumption}
\usepackage[utf8]{inputenc}
\usepackage{longtable}
\usepackage[ruled,vlined]{algorithm2e}
\usepackage{color}
\usepackage{caption}
\usepackage{multirow}
\usepackage[footnotesize]{subfigure}
\usepackage{hyperref}
\newtheorem{thm}{\textbf{Theorem}}
\newtheorem{lem}[thm]{Lemma}
\newtheorem{prop}{\textbf{Proposition}}

\begin{document}
	\title{  ISAC Super-Resolution Receiver via Lifted Atomic Norm Minimization}
	\author{Iman Valiulahi, Christos Masouros, \textit{Fellow,} IEEE, {\color{black} Athina P. Petropulu, \textit{Fellow}, IEEE} 
	\thanks{Tman Valiulahi and Christos Masouros are with the Department of Electronic and Electrical Engineering, University College London, London WC1E 7JE, U.K. (e-mails: i.valiulahi@ucl.ac.uk; c.masouros@ucl.ac.uk). {\color{black} Athina P. Petropulu is with the Department of
		Electrical and Computer Engineering, Rutgers, The State University of New
		Jersey, Piscataway, NJ 08854 USA (email:
		athinap@soe.rutgers.edu).} }	}
	
	\maketitle

	\begin{abstract}
		This paper introduces an off-the-grid estimator for integrated sensing and communication (ISAC) systems, utilizing lifted atomic norm minimization (LANM). The key challenge in this scenario is that neither the transmit signals nor the radar-and-communication channels are known. We prove that LANM can simultaneously achieve localization of radar targets and decoding of communication symbols, when the number of observations is proportional to the degrees of freedom in the ISAC systems. Despite the inherent ill-posed nature of the problem, we employ the lifting technique to initially encode the transmit signals. Then, we leverage the atomic norm to promote the structured low-rankness for the ISAC  channel. We utilize a dual technique to transform the LANM into an infinite-dimensional search over the signal domain. Subsequently, we use semidefinite relaxation (SDR) to implement the dual problem.
	We extend our approach to practical scenarios where received signals are contaminated by additive white Gaussian noise (AWGN) and jamming signals.  Furthermore, we derive the computational complexity of the proposed estimator and demonstrate that it is equivalent to the conventional pilot-aided ANM for estimating the channel parameters. Our simulation experiments demonstrate the ability of the proposed LANM approach to estimate both communication data and target parameters with a performance comparable to traditional radar-only super-resolution techniques.      
	\end{abstract}
	
	
	\section{Introduction}
	
	In recent years, there has been a significant interest in addressing the challenge of communication and radar spectrum sharing (CRSS) \cite{liu2020joint}.  In general, CRSS research is divided into two main directions, radar-communication coexistence and integrated sensing and communication (ISAC) system designs \cite{li2017joint,paul2016survey}.  First focuses on developing effective interference management techniques using a control center to coordinate two functions  without causing unwanted interference  \cite{vargas2023dual,li2016optimum,meng2024cooperative1}. ISAC techniques \cite{valiulahi2023net,alaaeldin2023robust}, however, concentrate on creating integrated systems capable of simultaneous performance of both wireless communication and remote sensing without the need of a control center. ISAC design offers advantages in both sensing and signaling operations through real-time cooperation.  This line of work has been expanded to various innovative applications, such as vehicular networks, indoor positioning, and covert communications \cite{wymeersch20175g,yang2015wifi,blunt2010intrapulse,hu2024isac}. 
	
ISAC systems aim to unify the operations of sensing and communication, seeking not only mutual performance enhancements but also a deeper integration paradigm where these functions are co-designed rather than perceived as distinct end-goals \cite{liu2022integrated, liu2018toward}. ISAC significantly enhances spectral and energy efficiencies while simultaneously reducing hardware and signaling costs \cite{meng2024cooperative} \cite{valiulahi2022antenna}. This is because of the fact that in the ISAC systems both sensing and communication operations are done with a unified system, in contrast with conventional communication and radar systems that exploit various resources. 

	\begin{figure*}
	\centering
	\includegraphics[width=1\linewidth]{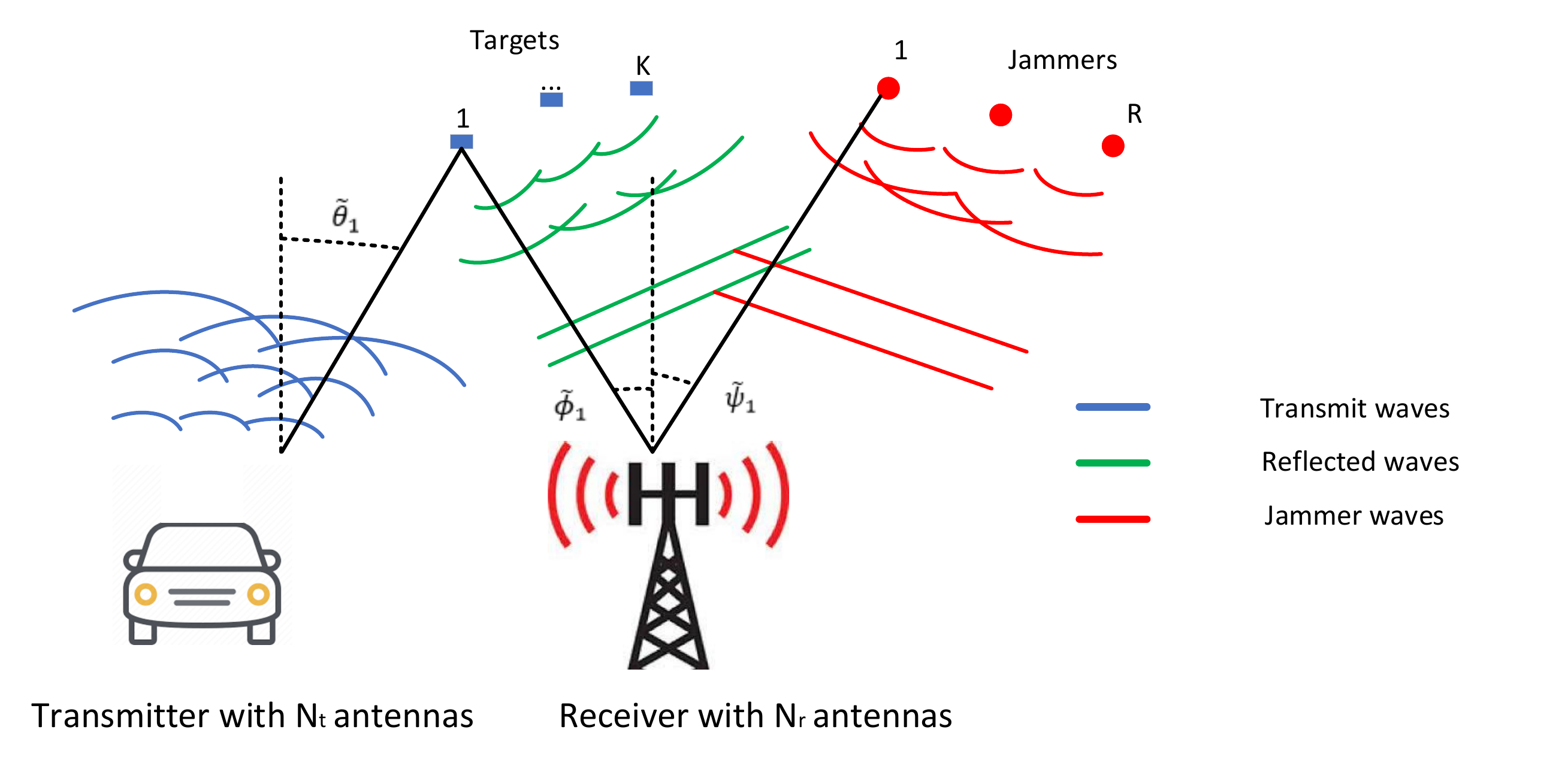}
	\caption{System model.}
	\label{fig:drawing1}
\end{figure*}

The difficult challenge in ISAC reception is that both the transmit signals and the radar response and communication channels are not known, which makes classical communications-only and radar-only approaches inapplicable. The main goal of ISAC receiver design is to concurrently estimate both the target parameters and the communication data of a remotely located transmitter from the echo signals reflected off the targets \cite{liu2020joint}. Accordingly, traditional approaches need to be adapted towards this goal. In terms of radar-only approaches, various methods have been proposed for traditional direction of arrival (DOA) estimation, including multiple signal classification (MUSIC) \cite{schmidt1986multiple} and estimating signal parameters via rotational invariance techniques (ESPRIT) \cite{yan2019two}. These methods utilize noise and signal subspaces, respectively. However, in highly correlated or coherent target scenarios, the performance of subspace decomposition methods is significantly diminished. These methods require prior knowledge of the number of targets, which is difficult to determine in advance in practice. Additionally, the performance of subspace decomposition methods is highly degraded in the presence of noise jamming. Pioneered by \cite{herman2009high}, compressed sensing (CS) emerged as a promising approach to solve the classical DOAs problem, reducing implementation complexity and improving detection resolution by utilizing the signal sparsity in the spatial domain. In \cite{yu2010mimo,rossi2012spatial}, the authors demonstrated that random antenna arrays can achieve nearly identical performance compared to the unitary linear array (ULA) implementation. However, the main assumption of CS for the DOA problem is that the DOAs are located on a fine grid, which contradicts practical scenarios where DOAs can be anywhere in the target area. Thus, a mismatch between inherent and estimated DOAs is undeniable, leading to a significant reduction in reconstruction performance \cite{chi2011sensitivity}.

To address the aforementioned basis mismatch, a novel off-the-grid CS approach, known as atomic norm minimization (ANM), has been developed to promote the sparsity of signals within a continuous dictionary \cite{tang2013compressed,valiulahi2019robustness}. More precisely, the ANM can be considered as a continuous version of the $\ell_{1}$ norm minimization, which is able to minimize the number of atoms required for the construction of the signal of interest over a continuous dictionary \cite{bigdeli2022noncoherent}. This method has found application in various communication and signal processing problems, including MIMO radar \cite{tang2020range}, line spectral estimation \cite{valiulahi2019two}, the elimination of impulsive noise in OFDM systems \cite{valiulahi2019eliminating}, and super-resolution problem \cite{valiulahi2019robustness}. For instance, in \cite{tang2020range}, the authors employed ANM to estimate the spatial characteristics of radar channels in a continuous domain, demonstrating its superior performance compared to conventional estimators. The ANM in \cite{heckel2016super} has been used to recover the continuous delay-Doppler pairs and corresponding attenuation
factors  under some mild minimum separation constraints. The work in \cite{heckel2016super1} also showed that the ANM can be used to detect the continuous angle-delay-Doppler
triplets and the corresponding attenuation factors.

On the other hand, a lifted technique has been exploited for blind detection in the super-resolution problem, where simultaneous estimation of both the sparse spikes in signals and the point spread function (PSF) is required \cite{chi2016guaranteed}, which is particularly relevant in fields like radar, sonar, and communications. By leveraging atomic norm minimization and lifting techniques, the authors in  \cite{chi2016guaranteed} provide performance guarantees under certain assumptions about the signal and the PSF. However, the method is limited to solving the one-dimensional problem and can only be applied to estimate delays. This makes it suitable for specific applications like radar and communication problems where delay estimation is critical, but not for more practical scenarios such as MIMO systems. The authors in \cite{suliman2021mathematical}  solved the blind super-resolution problem when the received signal is contaminated with additive white Gaussian noise (AWGN).  In \cite{vargas2023dual}, the authors used the LANM to distinguish the pulse radar signal from communication messages at the same bandwidth, in the context of communications-radar spectrum co-existence. 
 From a communication perspective, accurately estimating the transmit signal is crucial for decoding digital messages. 
 
 Consider a bistatic radar scenario where two access points are located at different positions compared to the expected target distance, as illustrated in Fig. \ref{fig:drawing1}. The first access point transmits signals, and the second one receives echoes from $K$ targets. The problem is to simultaneously detect the parameters of the targets and estimate the transmitted signals carrying communication data. Generally speaking, this is a bilinear problem, which is challenging to solve.  Traditionally, it is commonly assumed that there is a link for exchanging reference signaling between the transmitter and receiver or pilot signals, significantly wasting bandwidth.   The estimated signal might contain demodulation errors, leading to a significant reduction in estimation performance \cite{zheng2017super}. Moreover, a physical obstruction might block the radio waves from the transmitter to the receiver, consequently, the direct link might not be available. 
 
To address these issues, we propose a novel approach based on the lifted ANM (LANM) to concurrently estimate radar parameters and communication data. While the work in \cite{chi2016guaranteed} addresses the super-resolution problem for off-the-grid points, our approach utilizes LANM within the context of ISAC systems in a MIMO setting  to concurrently estimate four parameters—delay, Doppler shift, angle of arrival, and angle of departure. This approach is particularly significant for ISAC systems in $5$G and beyond, as it enhances the capability to manage complex, multi-dimensional environments using multiple antennas systems.
 Initially, we assume that the transmitted signals lie on a low-dimensional random dictionary, known by both the transmitter and receiver. Subsequently, we convert the received signal into a sparse combination of low-rank matrices, facilitating the identification of  the minimum number of low-rank matrices in the form of LANM. We propose the dual of LANM to construct the dual certificate, proving that our proposed problem can have an exact solution when the number of observations is proportional to the degrees of freedom and under certain mild conditions on the minimum separation of the targets.  Although the proposed dual problem is convex, it requires an infinite-dimensional search over the delay-Doppler and angular domains. To tackle this issue, we leverage the results of the trigonometric polynomial theory (TPT) \cite{dumitrescu2007positive} to develop linear matrix inequalities in terms of semidefinite relaxation (SDR). We then generalize our problem for practical scenarios where the received signal is contaminated by AWGN noise and jammer signals. Then, we investigate the computational complexity of the proposed estimator and show that it is equivalent to the pilot-aided ANM. This demonstrates that our approach can be applied to ISAC systems with the same implementation cost as the pilot-aided scenario. Finally, we conduct numerical simulations to evaluate the performance of our approach against state-of-the-art methods.

	Here, we introduce the notation used in this paper.
	Vectors and matrices are denoted by boldface lowercase and uppercase letters, respectively and scalars or entries are non-bold lowercase. The $\|\cdot\|_1$ and $\|\cdot\|_2$ are $\ell_1$ and $\ell_2$ norms, respectively. The operators $\mathrm{tr}(\cdot)$ and  $(\cdot)^H$ are trace of a matrix, hermitian of a vector, respectively.

	\section{System Model and Problem Formulation}
	As shown in Fig. \ref{fig:drawing1}, we consider an ISAC system with an $N_t$-antenna transmitter and an $N_r$-antenna receiver with $K$ targets, constrained on $K < N_{t}$. From Section \ref{sectionnoise} onwards, we further consider $R$ jammers that interfere with the ISAC signal. We assume that the targets are located at the far field of the arrays. The transmit
	and receive antennas are assumed to be uniformly spaced with distances $\frac{N_{t}}{2f_{c}}$
	and $\frac{N_{r}}{2fc}$
	, respectively, where $f_{c}$ is the carrier frequency. This leads to a uniformly spaced virtual array with $N_{t}N_{r}$ elements, which is the maximum number of achievable virtual antennas. The baseband received signal $y_{r}(t)$ by $N_{r}$ antennas
	$r = \{0, \cdots, N_{r}-1\}$  at the time index $t$ consists of the sum of the reflected signal from the targets of the transmitted
	probing signals, $x_{k}(t)$, $k = \{0, \cdots, K-1\}$, which can be given by
	\begin{align}\label{1}
		y_{r}(t)=\sum_{k=0}^{K-1}\sum_{s=0}^{N_{t}-1}\alpha_{k}e^{i2\pi r N_{t}\phi_{k}}e^{i2\pi s \theta_{k}}x_{k}(t-\bar{\tau}_{k})e^{i2\pi\bar{v}_{k}t},
	\end{align}
where $i=\sqrt{-1}$,  $\alpha_{k}=|\alpha_{k}|e^{i2\pi 
\theta_{k}^{\alpha}} \in \mathbb{C}$ is the complex attenuation factor, representing the effects of both the radar path-loss and cross-section for the $k$-th target,  $\theta_{k}=-\frac{\sin(\tilde{\theta}_{k})}{2},  \phi_{k}=-\frac{\sin(\tilde{\phi}_{k})}{2} \in [0,1]$ are angle of departure (AoD) and angle of arrival (AoA) of the $k$-th target, respectively, as shown in Fig. \ref{fig:drawing1}. $\bar{\tau}_{k}$ and $\bar{v}_{k}$ are the delay and Doppler shifts corresponding to the $k$-th target, respectively. The parameters $\theta_{k}, \phi_{k},  \bar{\tau}_{k}, \bar{v}_{k}$ can be translated to the
	angle, distance, and velocity of the $k$-th target relative to the radar. The aim of ISAC receiver design is to recover the parameters $\alpha_{k}, \theta_{k}, \phi_{k}, \bar{\tau}_{k}, \bar{v}_{k}$ from the received signal $y_{r}(t), r = \{0, \cdots, N_{r}-1\}$ and estimate the information-carrying  probing signals $x_{k}(t)$.   Note that there is no direct transmitter-receiver links assumed in traditional works. This link that can carry pilot signals, significantly wasting bandwidth and causing demodulation error \cite{zheng2017super}. In addition, a physical obstruction and building  might block the direct path. 	
	We assume that the transmit signal is band- and time- limited to be able to sample the received signal over finite indices. Let us assume that the probing signal $x_{k}(t)$ has bandwidth $B$ and contains a time interval proportional to $T$. According to (\ref{1}), one can understand that the band and time limitation assumptions on the probing signal imply that the received signal is
	band- and approximately time-limited, meaning that the delay-Doppler pairs are compactly
	supported. Thus, we can assume that the received signal is observed over the interval $[0,T_{t})$ and $(\bar{\tau}_{k}, \bar{v}_{k}) \in [0, T_{t}]\times[0, B] $.  Building on the fact that the received signal $y_{r}(t)$ is band-limited
	and approximately time-limited, we sample the received signals
	$y_{r}(t)$ over the interval $[0, T_{t}]$ at rate $\frac{1}{B}$, and collect the samples in a vector
	$\bm{y}_{r} \in \mathbb{C}^{\bar{L}}, \bar{L}:= BT$,  i.e., the $p$-th entry of $\bm{y}_{r}$ is $[\bm{y}_{r}]_{p} := y_{r}(\frac{p}{B})$, for $p = \{-N, \cdots,N\}, N = \frac{\bar{L}-1}{2}$, which can be given by
		\begin{align}\label{2}
		[\bm{y}_{r}]_{p}&=\frac{1}{\bar{L}}	\sum_{k=0}^{K-1}\sum_{s=0}^{N_{t}-1}|\alpha_{k}|e^{i2\pi r N_{t}\phi_{k}}e^{i2\pi (s \theta_{k}+\theta_{k}^{\alpha})}\nonumber\\&\sum_{r=-N}^{N}\Bigg[\Bigg(\sum_{l=-N}^{N}x_{k}(l)e^{-i2\pi\frac{rl}{\bar{L}}}\Bigg)e^{-i2\pi p\tau_{k}}\Bigg]e^{i2\pi\frac{rp}{\bar{L}}}\Bigg)e^{i2\pi v p},
	\end{align}
	where $\tau_{k}:=\frac{\bar{\tau}_{k}}{T}$ and $v_{k}:=\frac{\bar{v}_{k}}{B}$ are the normalized time- and frequency shifts, respectively, thus $(\tau_{k}, v_{k}) \in [0, 1)\times [0,1)$. Note that for obtaining (\ref{2}), it requires to first take Fourier
	transform, modulate the frequency, and take the inverse Fourier transform, which is a common approach for a continuous time-shift $\tau_{k}, v_{k}, \in [0,1)^{2}$ using the discrete probing signal $\bm{x}_{k}= [x_{k}(-N), \cdots, x_{k}(N)]^{T}$. The problem is now to identify the locations of the targets, which requires to estimate the parameters $\alpha_{k}$, ($\theta_{k}, \phi_{k}, \tau_{k}, v_{k}) \in [0,1)^{4}$, and simultaneously estimate probing signal $\bm{x}_{k}$, $ k = \{0, \cdots, K-1\}$ that contains communication information, from
the observation model in (\ref{2}).

	The order of unknown parameters of (\ref{2}) is $\mathcal{O}(\bar{L}S)$, which is much greater than the
	number of observations. Thus, the recovery problem is hopelessly
	ill-posed. To tackle this issue, we assume that the probing signals $\bm{x}_{k} \in \mathbb{C}^{\bar{L}\times 1}$,
	$\bm{x}_{k} := [x_{k}(-N), \cdots, x_{k}(N)]^{T}, k= \{0, \cdots, K-1\}$ lie on a known low-dimensional
	subspace defined by the known matrix $\bm{D} \in \mathbb{C}^{\bar{L}\times T}$
	with $T \ll \bar{L}$. Then, we can write $x_{k}=\bm{D}\bm{h}_{k}$	and $x_{k}(l) = \bm{d}_{l}^{H}\bm{h}_{k}$, where $\bm{h}_{k} \in \mathbb{C}^{T\times 1}, \forall k$ is the unknown orientation
	vector that carries communication data while $\bm{d}_{l} \in \mathbb{C}^{T\times 1}$ is the $l$-th column of $\bm{D}^{H}$, i.e,
	$\bm{D} := [\bm{d}_{-N}, \cdots, \bm{d}_{N}]^{H}$.  Without loss of generality, we assume
	that $\|\bm{h}_{k}\|_{2}= 1$ for all $k$. Note that recovering $\bm{x}_{k}$ is equivalent to
	estimating $\bm{h}_{k}$ since  $\bm{D}$ is known by both the transmitter and receiver.

	Now let us define $\bm{y}=[\bm{y}_{0}, \cdots, \bm{y}_{N_{r}-1}]^{T}$, $\bm{\tau}_{k}:=[\theta_{k}, \phi_{k}, \tau_{k}, v_{k}]^{T}$ and $\bm{a}(\bm{\tau}_{k})$ such that	
	\begin{align}\label{3}
	&	[\bm{a}(\bm{\tau}_{k})]_{(r, s, l, k, 1)}\nonumber\\&=e^{i2\pi rN_{t}\phi_{k}}e^{i2\pi (s\theta_{k}+\theta_{k}^{\alpha})}D_{N}\bigg(\frac{l}{L}-\tau_{k}\bigg)D_{N}\bigg(\frac{r}{L}-v_{k}\bigg), \nonumber\\
		& s \in \{0, \cdots, N_{t}-1\},~ r \in \{0, \cdots, N_{r}-1\}, \nonumber\\
		&  l,k \in \{-N, \cdots, N\},
	\end{align}
	where $D_{N}(t)$ is the Dirichlet kernel defined by $D_{N}(t) :=\frac{1}{\bar{L}}\sum_{m=-N}^{N}e^{i2\pi tm}$. Let us consider set $\mathcal{J}$ which considers all elements of $s, r, l, k$ such that $j= \{1, \cdots, L\}$ where $L:=\bar{L}N_{r}$. Starting from (\ref{2}), upon using (\ref{3}), and
	setting $x_{k}(l) = \bm{d}^{H}_{l}\bm{h}_{k}$, we can write
	\begin{align}
	[\bm{y}]_{j}=\sum_{k=0}^{K-1}|\alpha_{k}|[\bm{a}(\bm{\tau}_{k})]_{(r, s, l, k,1)}\bm{d}^{H}_{(j-l)}\bm{h}_{k}e^{i2\pi \frac{pr}{L}}.
	\end{align}
We now rewrite (\ref{2}) in a matrix-vector form. To do so,
	we first define 
	\begin{align}
		[\tilde{\bm{D}}_{j}]_{((k,l),1\to K)}=e^{i2\pi\frac{pr}{L}}\bm{d}_{j-l}^{H},~~ k,l \in \{-N, \cdots, N\}.
	\end{align}
	Building on the above definitions, we can write 
	\begin{align}
		[\bm{y}]_{j}&=\sum_{k =0}^{K-1}|\alpha_{k}|\bm{a}(\bm{\tau}_{k})^{H}\tilde{\bm{D}}_{j}\bm{h}_{k}=\mathrm{Tr}(\tilde{\bm{D}}_{j}\bm{U})=\langle\bm{U},\tilde{\bm{D}}_{j}^{H}\rangle,
	\end{align}
	where $\bm{U}=\sum_{k=0}^{K-1}|\alpha_{k}|\bm{h}_{k}\bm{a}(\bm{\tau}_{k})^{H}$. Let us define the linear operator $\mathcal{X} : \mathbb{C}^{K \times L^{2}} \to \mathbb{C}^{L}$ as
	\begin{align}
		[\mathcal{X}(\bm{U})]_{j}=\mathrm{Tr}(\tilde{\bm{D}}_{j}\bm{U}),~~j\in\{1, \cdots, L\},
	\end{align}
	and its adjoint as
	\begin{align}
	\mathcal{X}^{\star}(\bm{q})=\sum_{j=1}^{L}[\bm{q}]_{j}\tilde{\bm{D}}^{H}_{j}, ~\bm{q} \in \mathbb{C}^{L\times 1}.
	\end{align}
	Then, we can write
	\begin{align}\label{eq9}
		\bm{y}=\mathcal{X}(\bm{U}).
		\end{align}

 Equation
	(\ref{eq9}) implies that recovering the unknowns must be done using $\bm{U}$. It is worth noting that $\bm{U}$  is a sparse linear combination of multiple matrices over the following set of atoms 
	\begin{align}
	\mathcal{A} = \{ \bm{v}\bm{a}(\bm{\tau})^{H} \bigg| \bm{\tau} \in [0,1)^{4} , \|\bm{v}\|_{2} =1\}.
	\end{align}
We now define  the atomic norm  of matrix $\bm{U}$ as a
	gauge function corresponding to the convex hull of $\mathcal{A}$, 
	$\mathrm{conv}(\mathcal{A})$, as below
	\begin{align}
	\|\bm{U}\|_{\mathcal{A}}
	&= \mathrm{inf}\,\{t > 0:\bm{U} \in t\mathrm{conv} (\mathcal{A})\}, \nonumber\\
	&=  
	\inf_{\bm{\tau}_{k} \in [0,1)^{4}, \bm{v}_{k} \in {\mathbb{C}}^N, \|\bm{v}_{k}\|_{2} =1} \, \Big\{\sum_{k} \alpha_{k}\Big| \nonumber\\& \bm{U}=\sum_{k=0}^{K-1}\alpha_{k}\bm{h}_{k}\bm{a}(\bm{\tau}_{k})^{H}, \alpha_{k} \geq 0 \Big\}.
	\end{align}
To estimate $\mathcal{X}(\bm{U})$ from
	(\ref{eq9}), we propose the following LANM subject to the observation constraint as
	\begin{align}\label{problemmain}
	\min_{\bm{U}} \|\bm{U}\|_{\mathcal{A}},~~ \mathrm{s.t.}~~ \bm{y}=\mathcal{X}(\bm{U}).
	\end{align}
It is tough to solve (\ref{problemmain}) because the primal
	variable is infinite-dimensional over $\bm{\tau} \in [0,1)^{4}$. We propose the following dual of the LANM to tackle this issue
	\begin{align}\label{14}
		\max_{\bm{q}} &~\langle\bm{q},\bm{y}\rangle_{\mathbb{R}}\nonumber\\
		&\|\mathcal{X}^{\star}(\bm{q})\|_{\mathcal{A}}^{\star}\le 1,
	\end{align}
where $\bm{q}$ is the dual variable and $\|\cdot\|_{\mathcal{A}}^{\star}$ is the dual of lifted atomic norm. In the next section, we provide the main results of this paper.

\section{Conditions of Solvability of the LANM Problem}
We first introduce the main assumptions of the paper, then, we propose our main theorem.

\begin{assumption}\label{assum1}
 We assume that the columns of $\bm{D}^H$, i.e., $\bm{d}_{l} \in \mathbb{C}^{T\times 1}$, are independently
 and identically selected from a population $\mathcal{F}$ which satisfies the following isotropy and incoherence properties, respectively,
	\begin{align}
		&\mathbb{E}[\bm{d}_{l}\bm{d}_{l}^{H}]=\bm{I}_{T}, ~ l=\{-N, \cdots, N\}, \nonumber\\
		&\max|d(i)|^{2} \le \mu,
	\end{align}
where $\mu$ is the coherence parameter and $\bm{I}_{T}$ is an identical matrix with size $T\times T$. Moreover, we assume that $\mu T\geq 1$, for ease of theoretical analysis. This can be always ensured by selecting $\mu$ sufficiently large.
\end{assumption}

\begin{assumption}\label{assumption2}
We assume that the element vectors $\bm{h}_{k}$ for all $k$ are randomly selected over the complex unit sphere, i.e., $\|\bm{h}_{k}\|_{2} = 1$.
\end{assumption}

\begin{assumption}\label{assumption3}
The radar parameters,
$(\tau_{k}, v_{k}, \theta_{k}, \phi_{k})$, $k =\{ 1, \cdots, k\}$, require to satisfy the following separation conditions
\begin{align}
	\underset{\forall k. k^{\prime}, k\ne k^{\prime}}{ \min} \max \bigg(& |\tau_{k}-\tau_{k^{\prime}}|,|v_{k}-v_{k^{\prime}}|,\nonumber\\&|\theta_{k}-\theta_{k^{\prime}}|,|\phi_{k}-\phi_{k^{\prime}}|\bigg)\geq \frac{10}{N_{t}N_{r}-1}
\end{align}
where $|a-b|$ is the wrap-around distance on the unit circle, i.e., the distance between $0.1$ and $0.9$ is $0.2$.
\end{assumption}

Note that 
the assumptions that $\bm{D}$ and
$\bm{h}_{k}$ are random, which are required for our proof, do not seem to be crucial in practice. More precisely, $\bm{D}$ which is known at the transmitter and receiver, can be considered as a coding matrix and $\bm{h}_{k}$ can cover different quadrature amplitude modulation (QAM) signals to convey communication information.  It is worth noting that normalizing the norm of $\bm{h}_{k}$ can still carry information for different QAM constellations by considering the fact that the receiver knows which modulation scheme is used and the size of the transmit signal. Let us assume that the transmitter sends $\bm{h}_{1}= [1+j, 1-j]$ with $\|\bm{h}_{1}\|_{2}=2$, then at the receiver $\hat{\bm{h}}_{1}=[0.5+0.5j,   0.5-0.5j]$ can be detected. Regarding the fact that the norm of the transmit signal with the size of $2$ for the $4-$QAM is always $2$, one can recover the transmit symbols. This is more challenging for larger constellations. For example, let us assume that the transmitter exploits $16-$QAM and aims to transmit $\bm{h}_{1}=[1-3j, 1+3j]$. The recovered signal at the receiver is   $\hat{\bm{h}}_{1}=[0.2236-0.6708j, ~ 0.2236+ 0.6708j]$. Building on the fact that the receiver is aware of the size of the transmit signal and the constellation scheme, the norm of the transmit  signal can have three different values  as $2, 4.4721,~ \text{and}, ~6$. Thus, by multiplying $4.4721$ to the recovered signal, the receiver can detect the transmit symbols. Note that multiplying $2$ or $6$ leads to points where do not belong to the $16-$QAM constellation.    

	The minimum separation condition provided in Assumption \ref{assumption3} is necessary to avoid the radar parameters to be overlapped , resulting in a severely ill-posed
	problem. 	To elaborate more, let us explain a well-known one-dimensional super-resolution problem that aims to estimate the frequency parameters from a low-resolution measurement. The authors in \cite{tang2013compressed} show that as long as the minimum separation between the frequencies is less than $\frac{1}{N}$, where $N$ is the number of observations,  there is a pair of spike signals with the same minimum
	separation, hence no estimator can detect them. 
 Obviously, any separation constraint for such a simple problem is required for your problem that aims to estimate all parameters of MIMO radar scenario.
 Building on the above discussion, we are now ready to present our main theorem as follows:
 
 \begin{thm}\label{maintheorem}
 Consider the linear system in (\ref{1}) and its sampled
 version in (\ref{2}) and assume that the unknown waveforms vectors
 can be written as $\bm{x}_{k} = \bm{D}\bm{h}_{k}$ where $\bm{D}$ satisfies Assumption \ref{assum1}
 while $\bm{h}_{k}$ follows Assumption \ref{assumption2}. Further, let the unknown
 shifts satisfy the minimum separation in Assumption \ref{assumption3}. As long as, with probability at least $1-\delta$, 
 \begin{align}
 	L^{4} \geq C\mu KT \log\bigg(\frac{10KT}{\delta}\bigg),
 \end{align}
one can recover $\bm{U}$ using problem (\ref{problemmain}). The proof is given in the Appendix \ref{proofthepo}.
 \end{thm}

Theorem \ref{maintheorem}  shows that when the number of radar samples are more than a logarithmic factor, there is an exact solution for problem (\ref{maintheorem}) under  mild separation conditions on radar parameters. In the following section, we evaluate the optimality conditions of the proposed estimator and its solving approach.

	\section{Optimality Conditions and SDR}\label{optimality}
	In this section, we explain how the unknown radar parameters can be estimated through dual problem (\ref{14}).  Let us first start with the constraint of problem  (\ref{14}) 
	\begin{align}
	\|\mathcal{X}^{\star}(\bm{q})\|_{\mathcal{A}}^{\star}&=\underset{\bm{\tau}\in [0,1)^{4}, \|\bm{h}\|_{2}=1}{\sup}|\langle\bm{h},\mathcal{X}^{\star}(\bm{q})\bm{a}(\bm{\tau})\rangle|\nonumber\\
	&=\underset{\bm{\tau}\in [0,1)^{4}}{\sup}\|\mathcal{X}^{\star}(\bm{q})\bm{a}(\bm{\tau})\|_{2}\le 1.
	\end{align}
This implies that the first constraint of (\ref{14}) is  equivalent to bounding 
	the norm of the four-dimensional trigonometric vector polynomial $f (\bm{\tau}) =
	\mathcal{X}^{\star}
(\bm{q})\bm{a}(\bm{\tau})$  by $1$.  In the following proposition, we introduce the main features of the dual polynomial to identify the radar parameters
	upon using the dual solution. 
	
	\begin{prop}\label{firstpro}
		Let us assume that the atom set $\mathcal{A}$ contains atoms of the form $\bm{h}\bm{a}(\bm{\tau})^{H}$ constrained on $\|\bm{h}\|_{2}=1$ for $\bm{\tau} \in [0,1)^{4}$. Let us  define $\mathcal{R}=\{\bm{\tau}_{i}\}_{i=0}^{K-1}
	$ and consider $\bm{U}$ as the optimal solution of (\ref{maintheorem}). The optimal solution is unique as long as the
		following conditions are satisfied:

	1) There is a four-dimensional trigonometric vector polynomial $f(\bm{\tau})=
	\mathcal{X}^{\star}(\bm{q})\bm{a}(\bm{\tau})$ such that:
	\begin{align}\label{18}
		&f(\bm{\tau}_{k})=\frac{1}{\|\bm{h}_{k}\|_{2}}\mathrm{sign}(c_{k})\bm{h}_{k},~~~\forall \bm{\tau}_{k} \in \mathcal{R}\nonumber\\
		&\|f(\bm{\tau})\|_{2} <1, ~ \forall \bm{\tau}\in [0,1)^{4}\setminus \mathcal{R}.
	\end{align}

	2) $\Bigg\{\begin{bmatrix}
		\bm{a}(\bm{\tau}_{k})^{H}\tilde{\bm{D}}_{-N} & \\
		\cdots & \\
		\bm{a}(\bm{\tau}_{k})^{H}\tilde{\bm{D}}_{N}
	\end{bmatrix}\Bigg\}_{k=0}^{K-1}$
	is a linearly independent set. The proof of the proposition is given in Appendix \ref{app}.
\end{prop}

\begin{figure*}[t]
	\centering
	\mbox{
		\subfigure[]{\includegraphics[width=0.49\textwidth]{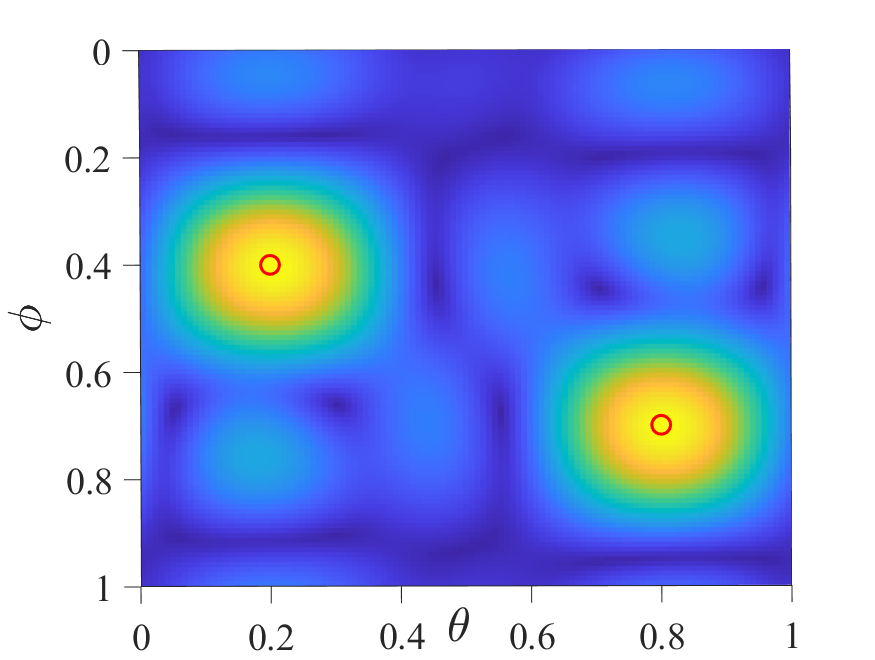}\label{fig.bound45}}
		\subfigure[]{\includegraphics[width=0.5\textwidth]{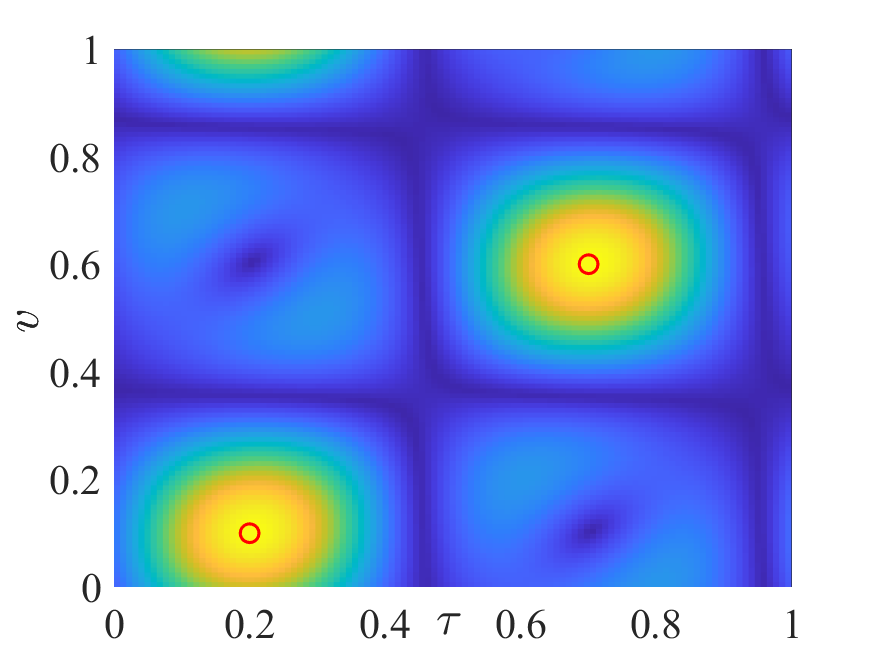}\label{fig.bound5}}
	}
	\caption{ The magnitude of the dual polynomial in $(\phi, \rho)$ and $(\tau, v)$ domains in Figs. \ref{fig.bound45} and \ref{fig.bound5}, respectively. Note that  red circles represent the recovered radar parameters, respectively.} \label{fig.phase1}
\end{figure*}

Once (\ref{14})  is solved and  $\bm{q}$ is obtained,
one can formulate $f(\bm{\bm{\tau}})$ in order to estimate $\mathcal{R}$ by discretizing the interval $[0, 1)^{4}$ on a fine grid and find $\|f(\hat{\bm{\tau}}_{k})\|_{2}=1$ regarding the fact that $\|\bm{h}_{k} \|_{2} = 1$ from Proposition \ref{prop1} (see Fig. \ref{fig.phase1}). Then, we propose the following least-square problem to estimate  $\alpha_{k}\bm{h}_{k}$. Let us first define  $\bm{g}_{k}=\alpha_{k}\bm{h}_{k}$, then solve
\begin{align}\label{12}
	\min_{\bm{g}_{k}, \forall k}\sqrt{\sum_{j=1}^{L}\bigg([\bm{y}]_{j}-\sum_{k =0}^{K-1}\bm{a}(\hat{\bm{\tau}}_{k})^{H}\tilde{\bm{D}}_{j}\bm{g}_{k}\bigg)^{2}}.
\end{align}
Building on the  fact that $\|\bm{h}_{k}\|_{2}=1$, one can obtain $\hat{\bm{h}}_{k}$ and $\hat{\alpha}_{k}$. 

It is worth noting that solving (\ref{14}) is still challenging because the inequality constraint requires an infinite dimensional search over the interval $[0, 1)^{4}$. We use the results of the TPT \cite{dumitrescu2007positive} to propose an SDR in terms of matrix inequalities. To do so, please let us first define sum-of-squares relaxation
degrees $s^{\prime}, r^{\prime}, l^{\prime}, k^{\prime}$. Then, we can define a zero-padded extension of $\bm{q}$
such that $\bm{q}_{(s^{\prime}, r^{\prime}, l^{\prime}, k^{\prime})}^{\prime}=\bm{q}_{s, r, l, k}$ if $s, r, l, k \in \mathcal{J}$, otherwise, $\bm{q}_{(s^{\prime}, r^{\prime}, l^{\prime}, k^{\prime})}^{\prime}=0$. Then, the SDR of problem (\ref{14}) can be given by
	\begin{align}\label{20}
		&\max_{\bm{q}, \bm{Q}\succeq 0}~ \langle \bm{q}, \bm{y} \rangle_{\mathbb{R}}\nonumber\\
		&\begin{bmatrix}
			\bm{Q} & \hat{\bm{Q}}^{H} \\
			\hat{\bm{Q}} &  \bm{I}_{T\times T}, 
		\end{bmatrix}\succeq 0,\nonumber\\
	&\mathrm{Tr}\bigg(\bigg(\bm{\Theta}_{l^{\prime}}\otimes\bm{\Theta}_{k^{\prime}}\otimes\bm{\Theta}_{s^{\prime}}\otimes\bm{\Theta}_{r^{\prime}}\bigg)\bm{Q}\bigg)=\delta_{l^{\prime}, k^{\prime}, s^{\prime}, r^{\prime}},
	\end{align}
where $-(N-1)\le l^{\prime}, k^{\prime} \le N-1$, $ -(S-1)\le s^{\prime} \le S-1$ and $ -(N_{r}-1)\le r^{\prime} \le N_{r}-1 $. Moreover, 
$\bm{\Theta}_{l^{\prime}} \in \mathbb{C}^{N\times N}$, $\bm{\Theta}_{k^{\prime}} \in \mathbb{C}^{N\times N}$, $\bm{\Theta}_{s^{\prime}} \in \mathbb{C}^{S\times S}$, $\bm{\Theta}_{r^{\prime}} \in \mathbb{C}^{N_{r}\times N_{r}}$ are Toeplitz
matrices with ones on their $l^{\prime}, k^{\prime}, s^{\prime}, r^{\prime}$-th diagonal and zeros elsewhere, respectively.
$\delta_{l^{\prime}, k^{\prime}, s^{\prime}, r^{\prime}}$ is the Dirac function such that $\delta_{l^{\prime}, k^{\prime}, s^{\prime}, r^{\prime}}=1$ if $l^{\prime}=k^{\prime}=s^{\prime}=r^{\prime}=0$, otherwise, $\delta_{l^{\prime}, k^{\prime}, s^{\prime}, r^{\prime}}=0$. Finally,
$
	\hat{\bm{Q}}=\sum_{j \in \mathcal{J}}q_{j}\bm{G}_{j}$, 
where $\bm{G}_{j}=\bm{e}_{j}\bm{d}_{l}, \forall l \in \{-N, \cdots, N\}$. 
Problem (\ref{20}) is convex and can be solved using off-the-shelf convex solvers such as CVX \cite{grant2014cvx}. In the next section, we generalize our problem for the case when the received signal is contaminated by AWGN noise and jamming signals.

\section{Robustness Against AWGN Noise and Jamming Signals}\label{sectionnoise}
In this section, we propose a robust LANM to show that the proposed estimator can work in the case when the received signal is contaminated by AWGN noise and jamming signals as shown in Fig. \ref{fig:drawing1}.  
To do so, we consider the following observation model 
\begin{align}
	\bm{y}_{w}=\bm{y}+\bm{w}+\bm{z},
\end{align}
containing both AWGN and jamming signals. 
The power of the AWGN is bounded by $\|\bm{w}\|_{2} \le \delta_{2}^{2}$ and the jamming signals have the following structure
\begin{align}
\bm{z}=\sum_{r=1}^{R}p_{r}\bm{p}_{r}\otimes \bm{a}_{N_{r}}(\psi_{r}),
\end{align}
where  $\bm{a}_{N_{r}}(\psi_{r})=[1, e^{j2\pi \psi} \cdots, e^{j2\pi \psi(N_{t}-1)} ]^{T}$ is the steering vector in which $\psi_{r}=-\frac{\sin(\tilde{\psi}_{k})}{2},  $  and $R$ is the number of jammers. Note that the received signal from a moving target, characterized by a particular angle of arrival and velocity concerning the airborne radar, undergoes spatial and temporal correlation \cite{richards2005fundamentals}. To elaborate, the array response for an individual target involves the Kronecker product of its spatial and Doppler steering vectors, adjusted for magnitude scaling. In contrast, a jammer exhibits spatial correlation but lacks temporal correlation from the radar's perspective \cite{ward1998space}. This results in the observed signal being a Kronecker product of the jammer's temporal samples and the spatial steering vector. 
\begin{figure}[t]
	\centering
	\mbox{
		\subfigure[]{\includegraphics[width=0.25\textwidth]{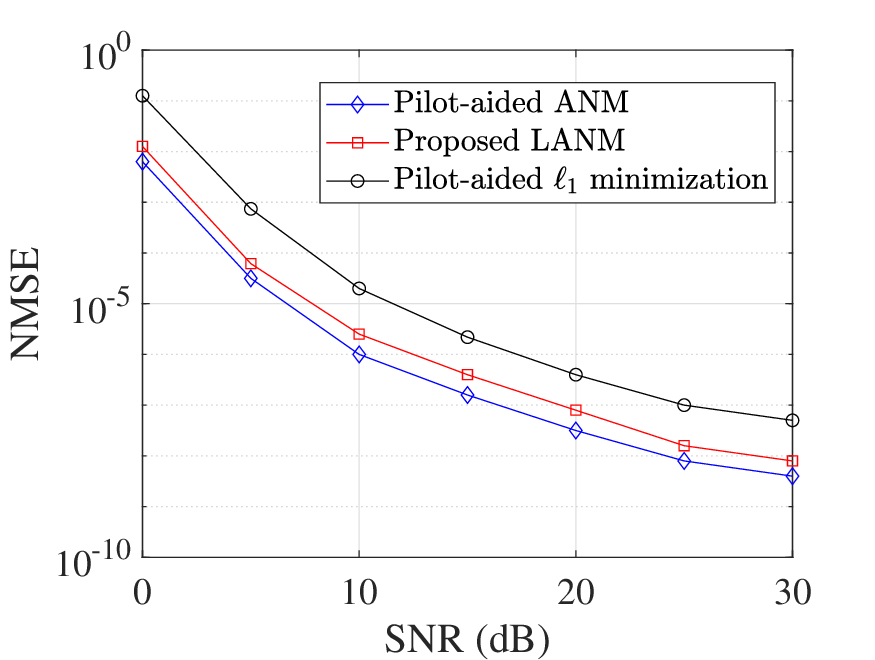}\label{fig.bound451}}
		\subfigure[]{\includegraphics[width=0.25\textwidth]{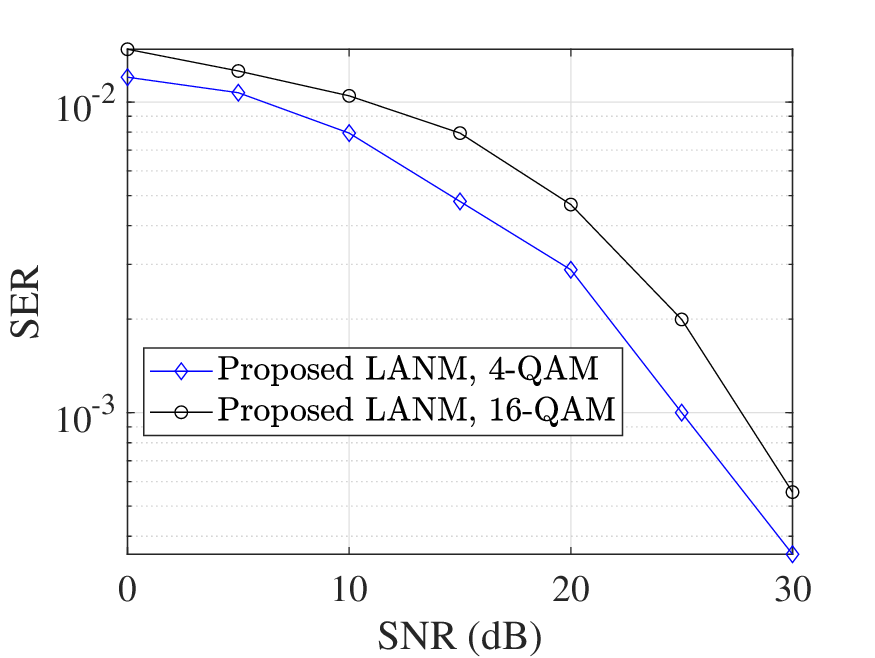}\label{fig.bound52}}
	}
	\caption{ The NMSE and SER of the proposed estimator compared to the conventional approaches in Figs. \ref{fig.bound451} and \ref{fig.bound52}, respectively. } \label{fig.phase2}
\end{figure}
\begin{figure*}[t]
	\centering
	\mbox{
		\subfigure[]{\includegraphics[width=0.3\textwidth]{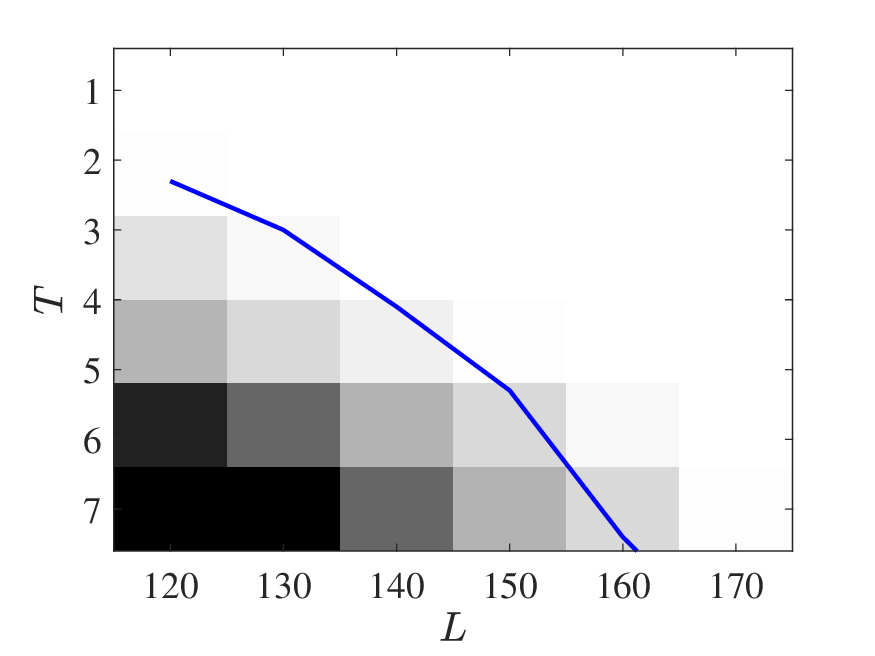}\label{fig.bound412}}
		\subfigure[]{\includegraphics[width=0.3\textwidth]{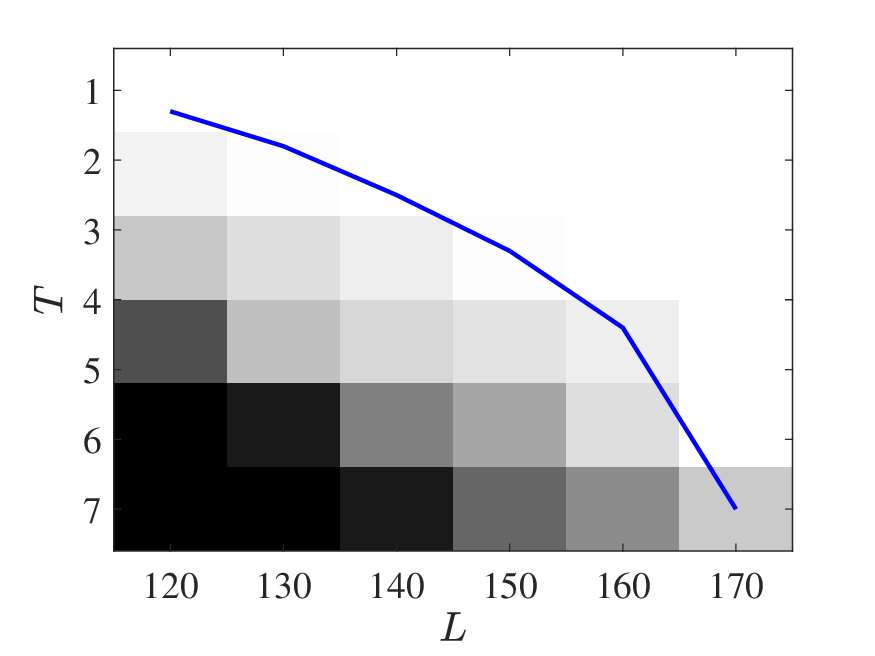}\label{fig.bound124}}
		\subfigure[]{\includegraphics[width=0.3\textwidth]{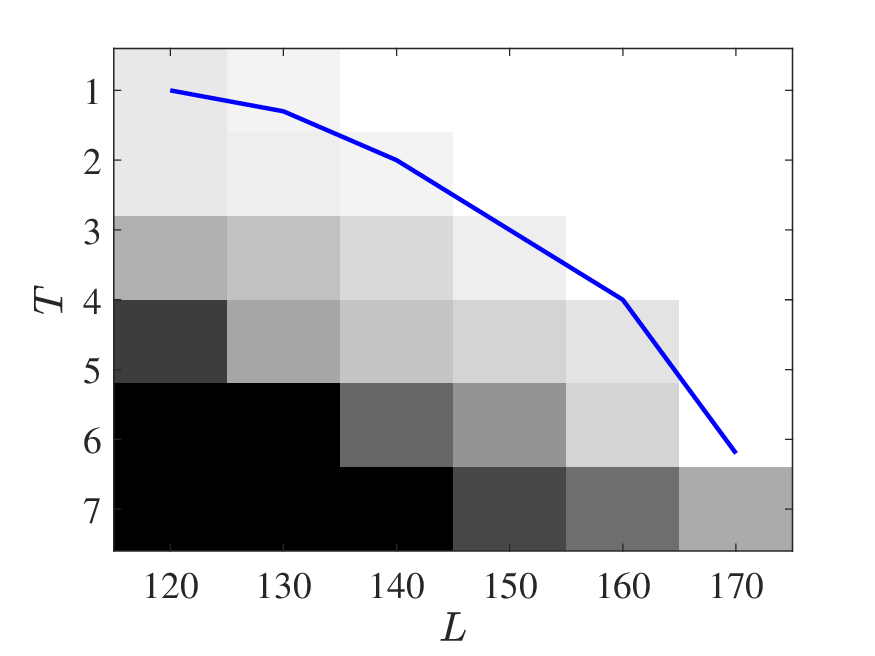}\label{fig.bound1245}}
		
	}
	\caption{The average success rate of LANM with respect to the number of targets  and the subspace dimension $T$ when
		the number of measurements varies. Figs. \ref{fig.bound412}, \ref{fig.bound412}, and \ref{fig.bound124} represent the results for $K=1$, $K=2$, $K=1$ with one jammer, respectively.  } \label{fig.phase4}
\end{figure*}
	\begin{figure}[b]
	\centering
	\includegraphics[width=.8\linewidth]{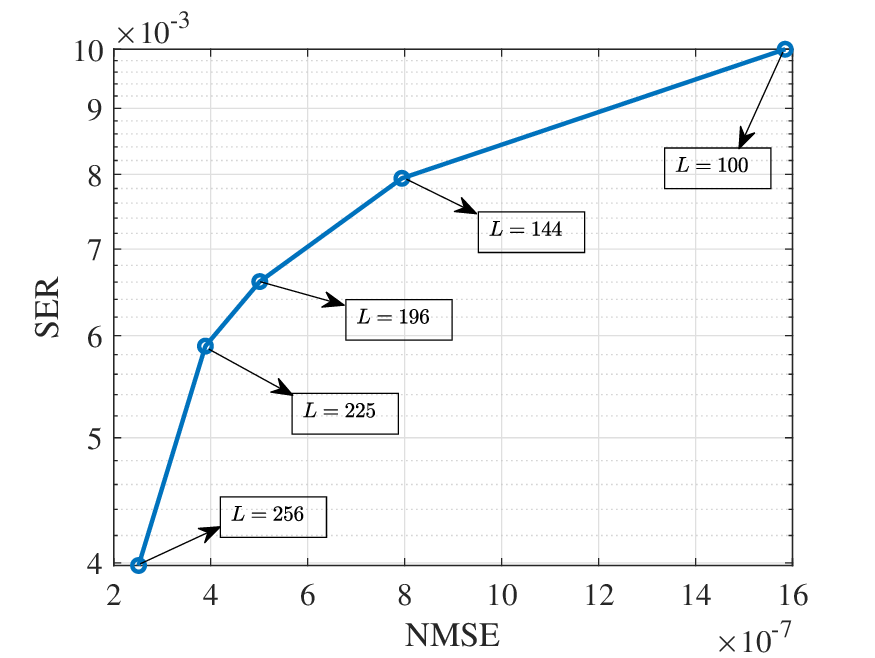}
	\caption{SER versus NMSE for different numbers of observations.}
	\label{fig:drawing12}
\end{figure}
Building on the fact that the jamming signals involve a sparse structure of the following atom set, 
\begin{align}
	\mathcal{A}_{j}=\bigg\{\bm{p}\otimes \bm{a}_{N_{r}}(\psi_{r}): \psi_{r} \in [0,1), \|\bm{p}\|=1\bigg\},
\end{align}
we define the following atomic norm for the jamming signals 
\begin{align}
	\|\bm{z}\|_{\mathcal{A}_{j}}= \inf \bigg\{\sum_{r}|p_{r}|: \bm{z}=\sum_{r}p_{r}\bm{p}_{r}\bm{a}_{N_{r}}(\psi_{r})\bigg\}.
\end{align}
Note that both the desired radar signal $\bm{y}$ and the jamming signal $\bm{x}_{r}$ are expressed as linear combinations of a small number of elements from the atomic sets $\mathcal{A}$ and $\mathcal{A}_{j}$ contaminated by AWGN noise. Consequently, we utilize the following atomic norm minimization to separate the signal components from $\bm{y}_{wr}$  as
\begin{align}
&\min_{\bm{U}, \bm{z}} \|\bm{U}\|_{\mathcal{A}}+\lambda \|\bm{z}\|_{\mathcal{A}_{j}}\nonumber\\
&\|\bm{y}_{w}-\bm{y}-\bm{z}\|_{2}\le 
\delta_{2}^{2},
\end{align}
where $\lambda >0$ is a regularization parameter that balances the sparsity of the lifted atomic norm of radar parameters and the sparsity of the jamming signals. The dual problem, then, can be given by 
	\begin{align}\label{27}
	\max_{\bm{q}} &~\langle\bm{q},\bm{y}_{w}\rangle_{\mathbb{R}}-\delta_{2}\|\bm{q}\|_{2}\nonumber\\
	&\|\mathcal{X}^{\star}(\bm{q})\|_{\mathcal{A}}^{\star}\le 1, \nonumber\\
	&\|\bm{q}\|_{\mathcal{A}_{j}}^{\star} \le \lambda,
\end{align}
 Following (\ref{20}), one can find a SDR for the above problem to implement it. Then, the transmitted data can be obtained through the following optimization 
\begin{align}\label{12}
	\min_{\bm{g}_{k}, \forall k, p_{r}\bm{p}_{r}, \forall r}\sqrt{\sum_{j=1}^{L} \bigg([\bm{y}_{w}]_{j}-\sum_{k =0}^{K-1}\bm{a}(\hat{\bm{\tau}}_{k})^{H}\tilde{\bm{D}}_{j}\bm{g}_{k}-[\hat{\bm{z}}]_{j})\bigg)^{2}}
\end{align}
where $\hat{\bm{z}} =\sum_{r=1}^{R}p_{r}\bm{p}_{r}\otimes \bm{a}_{N_{r}}(\hat{\psi_{r}})$.
To detect the location of the jammers, one can  use the locations of the norm of following polynomial  where achieve $\lambda$
\begin{align}
	\bm{Q}(\psi)= \bm{I}_{N_{r}} \otimes \bm{a}_{N_{r}}(\psi)\hat{\bm{q}},
\end{align} 
where $\hat{\bm{q}}$ is the solution of (\ref{27}) (see Fig. \ref{fig.bound457}). In the following section, we derive the computational complexity of the proposed approach.

\section{Complexity Analysis}
{\color{black} Problem \eqref{20} is convex and can be solved using the interior point method available by CVX \cite{grant2014cvx}. The computational complexity of solving such a problem is  $\mathcal{O}\left((E+F)^{1.5} E^2\right)$, where $E$ and $F$ represent the number of variables and constraints, respectively \cite{grant2014cvx}. In problem \eqref{20}, the number of variables is $(2N+1)N_{r}$ and $((2N+1)^2 N_{t} N_{r})^2$, and the number of constraints is $((2N+1)^2 N_{t} N_{r})^2 + 1$. To estimate the transmitted signal, we solve problem \eqref{12}, which is a mean-square problem. Since it has a closed-form solution, its complexity is negligible. Consequently, the overall computational complexity of the proposed estimator can be approximated as follows:
\begin{align}
	&\mathcal{O}\bigg((2N+1)N_{r}+((2N+1)^2 N_{t} N_{r})^2\nonumber\\
	&+((2N+1)^2 N_{t} N_{r})^2+1\bigg)^{1.5}\nonumber\\
	&\bigg((2N+1)N_{r}+((2N+1)^2 N_{t} N_{r})^2\bigg)^2 \nonumber\\
	&\approx \mathcal{O}\left(4N^2 N_{t} N_{r}\right)^3\left(4N^2 N_{t} N_{r}\right)^2 = \mathcal{O}\left(4N^2 N_{t} N_{r}\right)^5,
\end{align}
which shows that the number of receive antennas, $N_{r}$, transmit antennas, $N_{t}$, and the signaling dimension, $N$, significantly affect the computational complexity. The key insight of our approach is that this complexity is the same with the conventional atomic norm minimization to estimate the channel based on the pilot signal \cite{tang2020range}, as the complexity of estimating the transmit signal is neglected in problem \eqref{12} as there is a closed-form solution for it. However, due to the rapid increase in complexity with the number of antennas and signaling, one may develop an alternating direction method of multipliers (ADMM) for the proposed problem. This would provide a faster algorithm capable of handling more complex receiver and transmit systems efficiently.}

\section{NUMERICAL Results}
In this section, we explore the performance of the proposed LANM for the simultaneous estimation of radar parameters and transmit data. We compare the results with conventional approaches, including Atomic Norm Minimization (ANM) \cite{tang2020range} and $\ell_{1}$ minimization \cite{yu2010mimo}. The radar parameters are randomly generated, with coefficients for each target following i.i.d. zero-mean circularly symmetric complex Gaussian amplitude with unit variance. The number of observations and the size of the coding matrix will be specified in each simulation. For the jamming signal, we randomly generate the jammer's location and corresponding transmit signal. All ANM settings are the same as LANM except for the coding matrix. In the case of $\ell_{1}$ minimization, we discretize the continuous domain $[0,1)^{4}$. For the scenario where there are jammer in the target area, we assume that jammers transmits signals from a random location without overlapping with the targets. We consider problem (\ref{27}) by setting $\lambda=1$ for such a complex scenario. 
The communication signals are generated using QAM constellations, which will be specified for each simulation. The coding matrix, $\bm{D}$, follows a Gaussian distribution with zero mean and unit variance.

 Our evaluation begins by plotting the norm of the dual polynomial in (\ref{18}) in a simulation where the number of observed signals, $L=225$. Assuming two targets satisfying the minimum separation condition, locations where the dual polynomial's norm achieves $1$, as per Proposition \ref{firstpro}, are used to estimate radar parameters. In Fig. \ref{fig.phase1}, we plot the two-dimensional dual polynomial to illustrate this. The radar parameters, $(\tau_{k}, v_{k}, \theta_{k}, \phi_{k})$, are detected through the magnitude of the dual polynomial where achieve $1$.  For clarity, we present the dual polynomial in the two-dimensional case with two figures, plotting the pair $(\tau, v)$ for fixed $(\theta, \phi)$ and vice versa.

Before the next simulation, we need to first define the normalized mean-square error (NMSE)
as $\mathbb{E}[\|\bm{v}-\hat{\bm{v}}\|_{2}/\|\bm{v}\|_{2}]$, where $\bm{v}=\sum_{k=0}^{K-1}\alpha_{k}\bm{a}(\bm{\tau}_{k})^{H}$ and $\hat{\bm{v}}$ is the recovered radar quantity with the parameters $\hat{K}, \hat{\alpha}, \hat{\bm{\tau}}$. Next, we define the symbol error rate (SER) as $\mathrm{SER}=\mathbb{E}\big[\frac{1}{KT} \sum_{k=0}^{K-1}\sum_{t=1}^{T}\mathbb{I}_{(h_{k,t}-\hat{h}_{k,t})}\big]$
where $\mathbb{I}$ is a binary indicator such that $\mathbb{I}=0$  if $h_{k,t}=\hat{h}_{k,t}$, otherwise, $\mathbb{I}=1$.

We now compare the NMSE of LANM with the pilot-aided $\ell_{1}$ minimization, and ANM, as depicted in Fig. \ref{fig.bound451} versus signal-to-noise ratio (SNR) values, which is defined as $\text{SNR}=10\log_{10}\frac{1}{\delta_{2}}$.  In this simulation, we set  $L=225$. The figure illustrates that as the SNR thresholds increase, the NMSE decreases for all approaches. This observation highlights that, despite LANM simultaneously estimating the channel and transmit signal, its performance is comparable to the pilot-aided ANM and surpasses the traditional $\ell_{1}$-based method, both of which need knowledge of the transmit signal. It is worth noting that  because the transmitted signal is known at the receiver side in the case of pilot-aided ANM, its NMSE serves as a lower bound for the LANM. The performance of the LANM is superior to the pilot-aided $\ell_{1}$ minimization due to its increased robustness against additive noise and its ability to recover radar parameters within a continuous dictionary. This is shown in this figure that the performance of the proposed LANM is degraded when there is a jammer at the target area.  In Fig.  \ref{fig.bound52}, we investigate the SER of the proposed estimator for the different QAM constellations. This figure illustrates that SER decreases with increasing SNR thresholds. Moreover, Fig.  \ref{fig.bound52} indicates that the performance of the proposed estimator degrades by enlarging the QAM constellations.

\begin{figure*}[t]
	\centering
	\mbox{
		\subfigure[]{\includegraphics[width=0.495\textwidth]{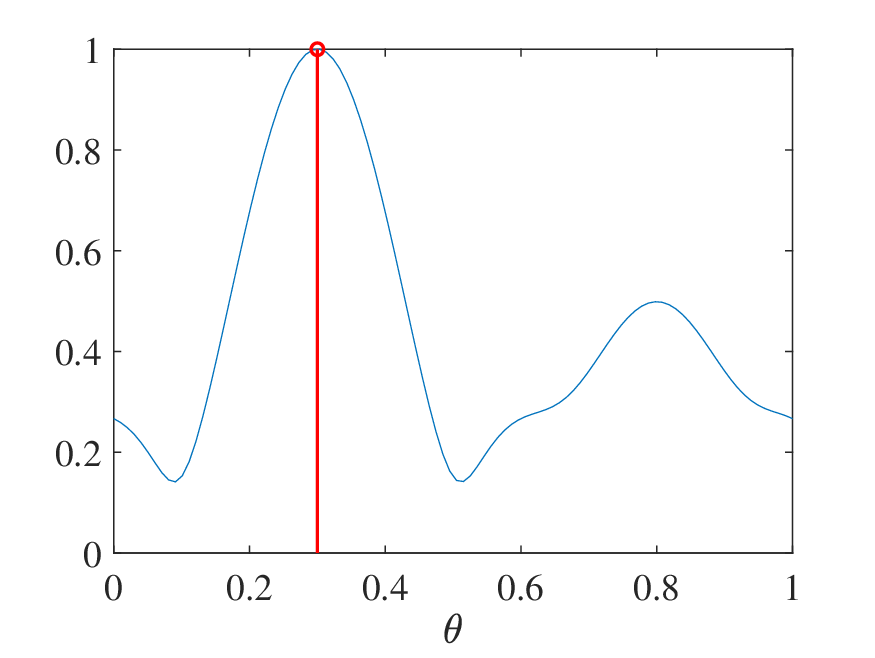}\label{fig.bound457}}
		\subfigure[]{\includegraphics[width=0.5\textwidth]{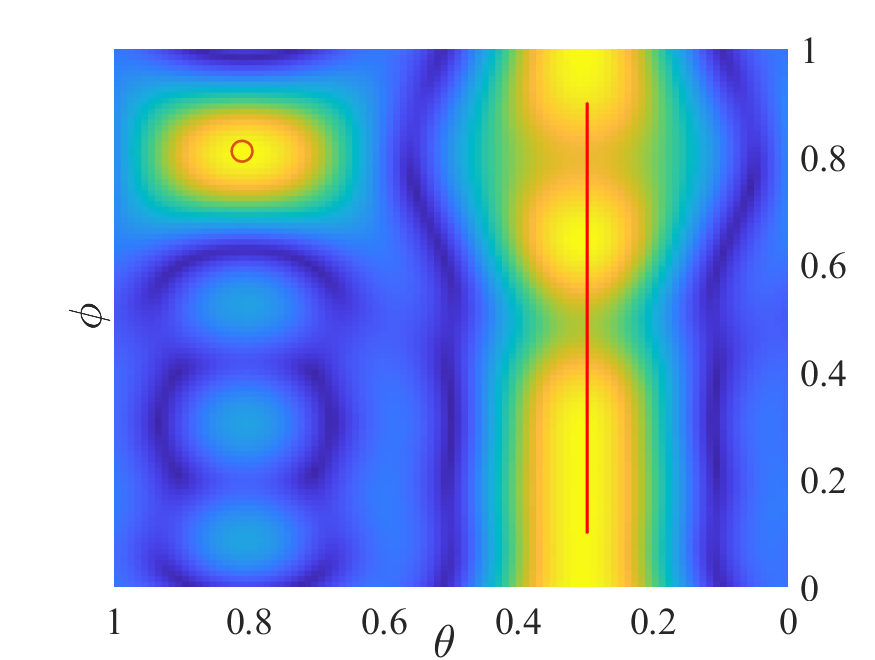}\label{fig.bound58}}		
	}
	\caption{ The dual polynomials for detecting the locations of the target and jammer. The red bar in Fig. \ref{fig.bound457} shows the jammer's parameter. The red circle and red bar demonstrate the recovered radar parameters and jammer interference, respectively.     } \label{fig.phase5}
\end{figure*}
 We proceed by examining the phase transition graphs of LANM with respect to the average success rate in relation to the number of targets and subspace dimension $T$ for various measurement quantities to show  the trade-off between the number of detected symbols and the recovery performance of radar channel. In each of the $20$ Monte Carlo simulations conducted, we calculate the normalized error $\|\bm{U} - \hat{\bm{U}}\|_{F}/\|\bm{U}\|_{F}$, where $\hat{\bm{U}}$ represents the recovered lifted matrix. If $\|\bm{U} - \hat{\bm{U}}\|_{F}/\|\bm{U}|_{F} \leq 10^{-3}$, we consider the recovery method successful. These figures reveal that an increase in the number of measurements correlates with an improvement in the number of detected symbols. Furthermore, a comparison between the results of Figs. \ref{fig.bound412} and \ref{fig.bound124} indicates that expanding the number of targets diminishes the recovery performance. These results are consistent with Theorem \ref{maintheorem}, where the recovery performance is a logarithmic function of the number of targets and the transmit data size. Moreover, in Fig. \ref{fig.bound1245}, we evaluate the phase transition graphs for the scenario where there is one target and one jammer in the target area. This figure demonstrates a logarithmic relationship between the number of targets and the transmit data size for this scenario, which we suggest eager readers to prove.

In the next phase of our study, we examine the radar-communication trade-off by varying the number of observed signals, as illustrated in Fig. \ref{fig:drawing12}. The figure indicates that an increase in the number of observations results in improvements in both the SER and NMSE of the radar channel. This observation is consistent with the notion that augmenting the number of observations enhances channel diversity, consequently leading to improved performance in both communication and radar metrics.

\begin{figure}[t]
	\centering
	\mbox{
		\subfigure[]{\includegraphics[width=0.25\textwidth]{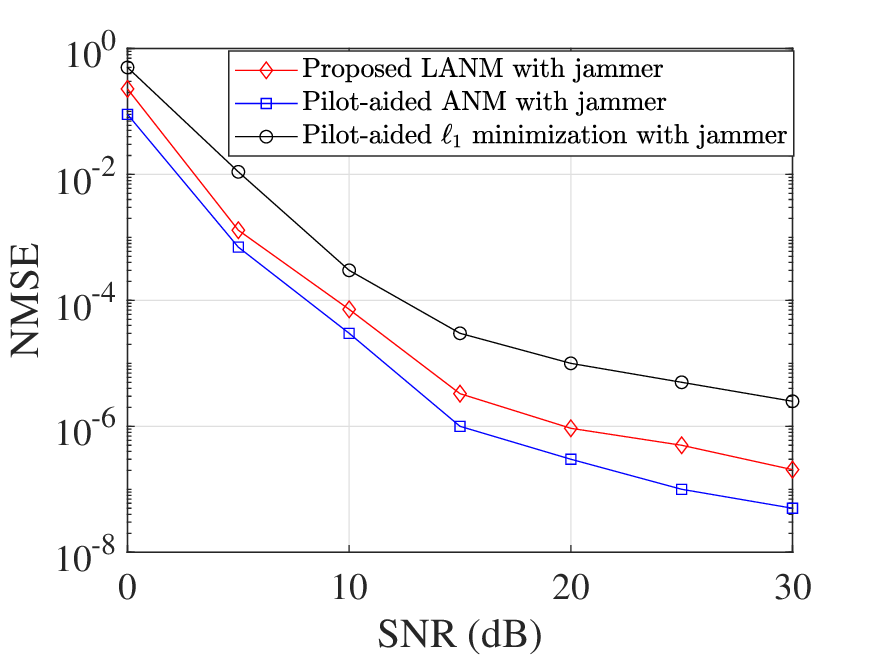}\label{fig.bound451jamer}}
		\subfigure[]{\includegraphics[width=0.25\textwidth]{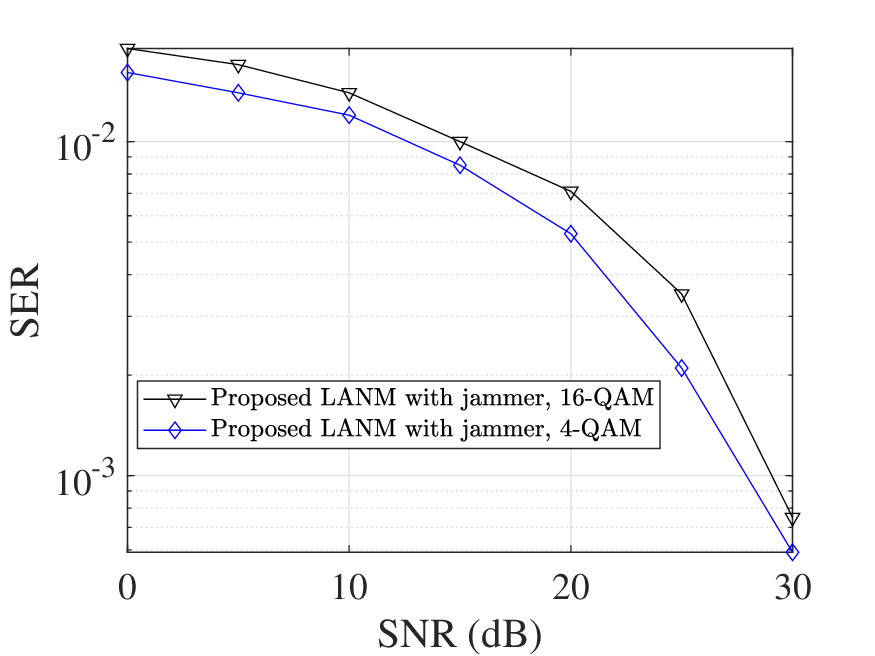}\label{fig.bound52jamer}}
	}
	\caption{ The NMSE and SER of the proposed estimator compared to the conventional approaches for $K=1$ with one jammer in the target area in Figs. \ref{fig.bound451jamer} and \ref{fig.bound52jamer}, respectively. } \label{fig.phase2jammer}
\end{figure}

For the next empirical finding, we plot the dual polynomials that contains jammer information and target's parameters. As shown in Fig. \ref{fig.bound457}. the jammer polynomial has one peak that shows the location of the jammer. Then, we plot the dual polynomial in the $(\phi, \theta)$  domain in Fig. \ref{fig.bound58}. This figure demonstrates that the dual polynomial has many peaks where the jammer is located and one peak that localizes the target. Note that we do not show the dual polynomial in the $(\tau, v)$ domain because the jammer has no interference with this plane. Finally,
We compare the NMSE of SER of the proposed estimator with the jammer with the pilot-aided $\ell_{1}$ minimization, and ANM. As shown in Figs. \ref{fig.bound451jamer} and \ref{fig.bound52jamer}, our proposed estimator outperform the conventional $\ell_{1}$ minimization in the NMSE and SER. Again, the performance obtained by the proposed estimator is comparable with the pilot-aided ANM as a performance benchmark.
	\section{Conclusion}
This paper presents a novel off-the-grid estimator for the ISAC systems, employing LANM. The study establishes that LANM can effectively accomplish both radar target parameter estimation and communication symbol decoding concurrently, subject to a mild constraint on the minimum distance of the targets and the number of observations. Despite the inherent challenges of the ill-posed problem, the paper adopts a lifting technique to encode transmit signals initially, utilizing atomic norm to enhance the structured low-rankness of ISAC systems. We then exploit the dual technique to transform the LANM into an infinite-dimensional search over the signal domain. Consequently, we employed SDR to implement the dual problem. Moreover, the paper extends the proposed approach to practical scenarios where received signals are contaminated by AWGN noise and jamming signals. The simulation experiments support the main theorem and demonstrate the superior performance of LANM  compared to the existing state-of-the-art methods.
	
\section{Appendices}
 \section{Proof of Theorem  \ref{maintheorem}}\label{proofthepo}
To prove Theorem \ref{maintheorem}, we first describe the
desired form of a valid  dual polynomial. Then, we show that this polynomial with its  optimal properties are satisfied, which concludes the proof.
\begin{prop}\label{prop1}
If the vector-valued dual polynomial
\begin{align}\label{validform}
	\bm{Q}(\bm{\tau})&=(\mathcal{X}^{\star}(\bm{q}))^{H}\bm{c}(\bm{\tau})=\bm{B}^{T}\mathrm{diag}(\bm{q})\bm{c}(\bm{\tau}),
\end{align}
satisfies
\begin{align}
	&\bm{Q}(\bm{\tau}_{k})=\frac{1}{\|\bm{h}_{k}\|_{2}^{2}}\mathrm{sign}(a_{k}^{\star})\bm{h}^{\star}_{k}, ~~~\forall \bm{\tau}_{k}\in \mathcal{T},\label{first}\\
	&\|\bm{Q}(\bm{\tau})\|_{2} <1, ~~~\forall \bm{\tau} \in [0,1)^{4} \setminus \mathcal{T}, \label{second}
\end{align}
where a
vector $\bm{q}$ is a vector and $\mathrm{sign}(\cdot)$ is the complex sign function, 	problem (\ref{problemmain}) has a unique solution. 
Using the above proposition, one can understand that $\bm{U}^{\star}$ is the unique and exact solution of problem (\ref{problemmain}) if one finds the dual polynomial $\bm{Q}(\bm{\tau})$, which satisfies (\ref{first}) and (\ref{second}). Thus, our main goal is to construct such a valid dual polynomial.
\end{prop}

\subsection{Construction of the dual polynomial}
Let us first suppose that  $\|\bm{h}_{k}\|_{2}=1, \forall k$, without loss of generality. Then, we can introduce the squared Fejer’s kernel and its partial derivatives as
\begin{align}
	K(\bm{\tau})&=\frac{1}{L^{4}}\sum_{\bm{n}\in \mathcal{J}}s_{n_{1}}s_{n_{2}}s_{n_{3}}s_{n_{4}}e^{-j2\pi \bm{\tau}^{T} \bm{n}},\nonumber\\
	&K^{i_{1}, i_{2}, i_{3}, i_{4}} (\bm{\tau})=\frac{\partial^{i_{1}}\partial^{i_{2}}\partial^{i_{3}}\partial^{i_{4}}K(\bm{\tau})}{\partial^{i_{1}}_{\tau_{1}}\partial^{i_{2}}_{\tau_{2}}\partial^{i_{3}}_{\tau_{3}}\partial^{i_{4}}_{\tau_{4}}},
\end{align}
where $s_{n}=\frac{1}{L}\sum_{i=\max(n-L,-L)}^{\min(n+L,L)}(1-|\frac{i}{L}|)(1-|\frac{n-i}{L}|)$
following the definition \cite{candes2014towards}. Now, we define the randomized
matrix-valued versions of $K(\bm{\tau})$ as
\begin{align}
	\bm{K}(\bm{\tau})=\frac{1}{L^{4}}\sum_{\bm{n}\in \mathcal{J}}s_{n_{1}}s_{n_{2}}s_{n_{3}}s_{n_{4}}\bm{d}_{\bm{n}}\bm{d}_{\bm{n}}^{H}e^{-j2\pi\bm{\tau}^{T} \bm{n}} \in \mathbb{C}^{L\times L}.
\end{align}
Similarly, the derivatives can be obtained. The expected value can be given by 
\begin{align}
	\mathbb{E}\{\bm{K}(\bm{\tau})\}&=\frac{1}{L^{4}}\sum_{\bm{n}\in \mathcal{J}}s_{n_{1}}s_{n_{2}}s_{n_{3}}s_{n_{4}}\mathbb{E}\{\bm{d}_{\bm{n}}\bm{d}_{\bm{n}}^{H}\}e^{-j2\pi \bm{\tau}^{T} \bm{n}}\nonumber\\&=K(\bm{\tau})\bm{I}_{L}.
\end{align}
Consequently,  $\mathbb{E}\{\bm{K}^{\prime}(\bm{\tau})\}=K^{\prime}(\bm{\tau})\bm{I}_{L}$, where $\bm{I}_{L}$ is the $L\times L$ identity matrix which satisfies the isotropy property. Now, we suggest the following form to construct the
vector-valued dual polynomial $\bm{Q}(\bm{\tau}) \in \mathbb{C}^{L}$ as 
\begin{align}
\bm{Q}(\bm{\tau})&=\sum_{k=1}^{K}\bm{K}(\bm{\tau}-\bm{\tau}_{k})\bm{\alpha}_{k}+\sum_{k=1}^{K}\bm{K}^{1000}(\bm{\tau}-\bm{\tau}_{k})\bm{\beta}_{k}\nonumber\\
&+\sum_{k=1}^{K}\bm{K}^{0100}(\bm{\tau}-\bm{\tau}_{k})\bm{\gamma}_{k}+\sum_{k=1}^{K}\bm{K}^{0010}(\bm{\tau}-\bm{\tau}_{k})\bm{\varphi}_{k}\nonumber\\&
+\sum_{k=1}^{K}\bm{K}^{0001}(\bm{\tau}-\bm{\tau}_{k})\bm{\varrho}_{k},
\end{align}
where $\bm{\alpha}_{k}=[\alpha_{k,1}, \cdots, \alpha_{k, L}]^{T} \in \mathbb{C} ^{L\times 1}$
 and
 $\bm{\beta}_{k}=[\beta_{k,1}, \cdots, \beta_{k, L}]^{T} \in \mathbb{C}^{L\times 1 }$,  $\bm{\beta}_{k}=[\beta_{k,1}, \cdots, \beta_{k, L}]^{T} \in \mathbb{C}^{L\times 1 }$,   $\bm{\gamma}_{k}=[\gamma_{k,1}, \cdots, \gamma_{k, L}]^{T} \in \mathbb{C}^{L\times 1 }$,   $\bm{\varphi}_{k}=[\varphi_{k,1}, \cdots, \varphi_{k, L}]^{T} \in \mathbb{C}^{L\times 1 }$, and $\bm{\varrho}_{k}=[\varrho_{k,1}, \cdots, \varrho_{k, L}]^{T} \in \mathbb{C}^{L\times 1 }$
  for $k= \{1, \cdots, K\}$. It is straightforward
to check that $\bm{Q}(\bm{\tau})$ follows the valid form defined in (\ref{validform}) of
Proposition \ref{prop1}. We form the following linear equations to find the coefficient vectors 
\begin{align}
	\bm{Q}(\bm{\tau}_{k})&=\mathrm{sign}(a_{k}^{\star})\bm{h}^{\star}_{k}, ~~\bm{\tau}_{k} \in \mathcal{T},\label{34} \\
	\bm{Q}^{\prime}(\bm{\tau}_{k})&=0, \hspace{1.7cm} \bm{\tau}_{k} \in \mathcal{T},\label{35}
\end{align}
which can be given by (\ref{eq21}) by defining $\kappa:=\frac{1}{\sqrt{K^{\prime\prime}(0)}}$ and $(\bm{\Gamma}_{i_{1}, i_{2}, i_{3}, i_{4}})_{lj}:=\bm{K}^{i_{1}, i_{2}, i_{3}, i_{4}}(\bm{\tau}_{l}-\bm{\tau}_{j})$. To solve this linear problem, we need to first show that $\bm{\Gamma}$ is invertible with high probability.
	\begin{figure*}
		\begin{align}\label{eq21}
			\begin{aligned}
			 \begin{bmatrix} 
			\bm{\Gamma}_{0000}	& \kappa \bm{\Gamma}_{1000}& \kappa \bm{\Gamma}_{0100} & \kappa \bm{\Gamma}_{0010} &  \kappa \bm{\Gamma}_{0001}  \\
				\bm{\Gamma}_{1000}	& -\kappa^{2} \bm{\Gamma}_{2000}& -\kappa^{2} \bm{\Gamma}_{110} & -\kappa^{2} \bm{\Gamma}_{1010} & -\kappa^{2} \bm{\Gamma}_{1001} \\
				\bm{\Gamma}_{0100}	& -\kappa^{2} \bm{\Gamma}_{1100}& -\kappa^{2} \bm{\Gamma}_{0200} & -\kappa^{2} \bm{\Gamma}_{0110} & -\kappa^{2} \bm{\Gamma}_{0101} \\
		\bm{\Gamma}_{0010}	& -\kappa^{2} \bm{\Gamma}_{1010}& -\kappa^{2} \bm{\Gamma}_{0110} & -\kappa^{2} \bm{\Gamma}_{0020} & -\kappa^{2} \bm{\Gamma}_{0011} \\
				\bm{\Gamma}_{0001}	& -\kappa^{2} \bm{\Gamma}_{1001}& -\kappa^{2} \bm{\Gamma}_{0101} & -\kappa^{2} \bm{\Gamma}_{0011} & -\kappa^{2} \bm{\Gamma}_{0002} \\
				\end{bmatrix}
			\begin{bmatrix}
				\bm{\alpha} \\
				\bm{\beta} \\
				\bm{\varphi}\\
				\bm{\varrho}\\
			\end{bmatrix}=
			\begin{bmatrix}
		\mathrm{sign}(\bm{a}_{1}^{\star})\bm{h}^{\star} \\
			\cdot \\
			\mathrm{sign}(\bm{a}_{K}^{\star})\bm{h}^{\star}\\
			0\\
			\cdot \\
		0
		\end{bmatrix} \longrightarrow \bm{\Gamma}	\begin{bmatrix}
		\bm{\alpha} \\
		\bm{\beta} \\
		\bm{\varphi}\\
		\bm{\varrho}\\
	\end{bmatrix}=
\begin{bmatrix}
\mathrm{sign}(\bm{a}_{1}^{\star})\bm{h}^{\star} \\
\cdot \\
\mathrm{sign}(\bm{a}_{K}^{\star})\bm{h}^{\star}\\
0\\
\cdot \\
0
\end{bmatrix},
			\end{aligned}
		\end{align}
	\end{figure*}

	\begin{figure*}
	\begin{align}\label{phi}
			\begin{aligned}
			\bm{\Phi}=\begin{bmatrix} 
				\bm{\Phi}_{0000}	& \kappa \bm{\Phi}_{1000}& \kappa \bm{\Phi}_{0100} & \kappa \bm{\Gamma}_{0010} &  \kappa \bm{\Phi}_{0001}  \\
				\bm{\Phi}_{1000}	& -\kappa^{2} \bm{\Phi}_{2000}& -\kappa^{2} \bm{\Phi}_{110} & -\kappa^{2} \bm{\Theta}_{1010} & -\kappa^{2} \bm{\Phi}_{1001} \\
				\bm{\Phi}_{0100}	& -\kappa^{2} \bm{\Phi}_{1100}& -\kappa^{2} \bm{\Phi}_{0200} & -\kappa^{2} \bm{\Theta}_{0110} & -\kappa^{2} \bm{\Phi}_{0101} \\
				\bm{\Phi}_{0010}	& -\kappa^{2} \bm{\Phi}_{1010}& -\kappa^{2} \bm{\Phi}_{0110} & -\kappa^{2} \bm{\Phi}_{0020} & -\kappa^{2} \bm{\Phi}_{0011} \\
				\bm{\Phi}_{0001}	& -\kappa^{2} \bm{\Phi}_{1001}& -\kappa^{2} \bm{\Phi}_{0101} & -\kappa^{2} \bm{\Phi}_{0011} & -\kappa^{2} \bm{\Phi}_{0002} \\
			\end{bmatrix}
	=\frac{1}{L^{4}}\sum_{\bm{n}\in \mathcal{J}}s_{n_{1}}s_{n_{2}}s_{n_{3}}s_{n_{4}}\bm{v}_{n}\bm{v}_{n}^{H},
			\end{aligned}
	\end{align}
	\end{figure*}
	The expectation value of $\bm{\Gamma}$ can be written as $\mathbb{E} \{\bm{\Gamma}\}=\bar{\bm{\Gamma}}=\bm{\Phi}\otimes \bm{I}_{L}$, where $\otimes$ is the Kronecker product, $\bm{\Phi} \in \mathbb{C}^{2K\times 2K}$ is given in (\ref{phi}), where $(\bm{\Phi}_{i_{1}, i_{2}, i_{3}, i_{4}})_{lj}=K^{i_{1}, i_{2}, i_{3}, i_{4}}(\bm{\tau}_{l}-\bm{\tau}_{j})$ and
	\begin{align}
		\bm{v}_{\bm{n}}&= [e^{-j2\pi \bm{\tau}_{1}^{T}\bm{n}},  \cdots, e^{-j2\pi \bm{\tau_{K}}^{T}\bm{n}}, \\
		& (j2\pi n_{1} \kappa)e^{-j2\pi \bm{\tau}_{1}\bm{n}}, \cdots, (j2\pi n_{1} \kappa)e^{-j2\pi \bm{\tau}_{K}^{T}\bm{n}}, \nonumber\\ &(j2\pi n_{2} \kappa)e^{-j2\pi \bm{\tau}_{1}^{T}\bm{n}}, \cdots, (j2\pi n_{2} \kappa)e^{-j2\pi \bm{\tau}_{k}^{T}\bm{n}}, \nonumber\\& (j2\pi n_{3} \kappa)e^{-j2\pi \bm{\tau}_{K}^{T}\bm{n}}, \cdots, (j2\pi n_{3} \kappa)e^{-j2\pi \bm{\tau}_{K}^{T}\bm{n}}, \nonumber\\
		&(j2\pi n_{4} \kappa)e^{-j2\pi \bm{\tau}_{K}^{T}\bm{n}}, \cdots, (j2\pi n_{4} \kappa)e^{-j2\pi \bm{\tau}_{K}^{T}\bm{n}}]^{T}
	\end{align}
	The following lemma is useful from \cite{candes2014towards} regarding $\bm{\Phi}$.

	\begin{lem}\label{lem}
 The matrix $\bm{\Phi}$ is
	invertible and 
	\begin{align}
		\|\bm{I}-\bm{\Phi}\| \le 0.1811,~ \|\bm{\Phi}\|\le 1.1811,~ \|\bm{\Phi}^{-1}\|\le 1.2301.
	\end{align}
	\end{lem}
To avoid excessive clutter, we eliminate the proof of 	Lemma \ref{lem}. However,  eager readers can \cite{candes2014towards} follow the process   to prove this lemma and obtain these bounds. It is worth noting that  $\bm{\Gamma}$ can be written as a sum of independent random
	matrices as
	\begin{align}
		\bm{\Gamma}&=\frac{1}{L^{4}}\sum_{\bm{n}\in \mathcal{J}}s_{n_{1}} s_{n_{2}}s_{n_{3}}s_{n_{4}}(\bm{v}_{\bm{n}}\otimes \bm{d}_{\bm{n}}) (\bm{v}_{\bm{n}}\otimes \bm{d}_{\bm{n}})^{H} \nonumber\\
		&=\frac{1}{L^{4}}\sum_{\bm{n}\in \mathcal{J}}s_{n_{1}} s_{n_{2}}s_{n_{3}}s_{n_{4}}(\bm{v}_{\bm{n}}\bm{v}_{\bm{n}}^{H})\otimes (\bm{d}_{\bm{n}}\bm{d}_{\bm{n}}^{H}),
	\end{align}
then, we can write $\bm{\Gamma}-\bar{\bm{\Gamma}}$ as  $\bm{\Gamma}-\bar{\bm{\Gamma}}=\sum_{\bm{n}\in \mathcal{J}}\bm{S}_{\bm{n}}$, where $\bm{S}_{\bm{n}}=\frac{1}{L^{4}}s_{n_{1}} s_{n_{2}}s_{n_{3}}s_{n_{4}}(\bm{v}_{\bm{n}}\bm{v}_{\bm{n}}^{H})\otimes (\bm{d}_{\bm{n}}\bm{d}_{\bm{n}}^{H}-\bm{I}_{L})$. Using the following lemma, we prove that $\bm{\Gamma}$ is concentrated around $\bar{\bm{\Gamma}}$ with high probability.

	\begin{lem}\label{lem2}
	Let us assume that $0 < \delta <1$, for any $\epsilon \in (0, 0.8189)$, if
	\begin{align}\label{mear}
		L^{4} \geq \frac{192 \mu KT}{\epsilon}\log\bigg(\frac{10KT}{\delta}\bigg),
	\end{align}
 $\|\bm{\Gamma}-\bar{\bm{\Gamma}}\|\le \epsilon$ holds with probability at least $1-\delta$. Proof of Lemma \ref{lem2} is given in Appendix \ref{appendix1}. 
	\end{lem}
	
	Note that the event $\mathcal{E}_{1, \epsilon}=\{\|\bm{\Gamma}-\bar{\bm{\Gamma}}\| \le \epsilon\}$ holds if  (\ref{mear}) holds 
	with probability at least $1-\delta$. This lemma states that the matrix $\bm{\Gamma}$ is invertible when $\mathcal{E}_{1, \epsilon}$ holds for
	some $0 < \epsilon < 0.8189$, because
\begin{align}
	\|\bm{I}-\bm{\Gamma}\| \le \|\bm{I}-\bar{\bm{\Gamma}}\|+\|\bar{\bm{\Gamma}}-\bm{\Gamma}\|\le 0.1811+\epsilon <1,
\end{align}
	where $\|\bm{I}-\bar{\bm{\Gamma}}\|=\|\bm{I}-\bm{\Phi}\|\le 0.1811$ from Lemma \ref{lem}.
		 Under $\mathcal{E}_{1, \epsilon}$, 
	 	 let $\bm{\Gamma}^{-1}=[\bm{L} \bm{R}]$, where 
	 	 $\bm{L}\in \mathbb{C}^{2LK\times LK}$ and
	 	  $\bm{R}  \in \mathbb{C}^{2LK\times LK}$, then we have 
	\begin{align}
		&[\bm{\alpha}, \kappa^{-1}\bm{\beta}, \kappa^{-1}\bm{\gamma}, \kappa^{-1}\bm{\varphi}, \kappa^{-1}\bm{\varrho}]^{T}\nonumber\\
		&=\bm{\Gamma}^{-1}[\mathrm{sign}(a_{1}^{\star})\bm{h}^{\star}, \cdots, \mathrm{sign}(a_{K}^{\star})\bm{h}^{\star}, 0, \cdots, 0]^{T}\nonumber\\
		&=\bm{L}[\mathrm{sign}(\bm{a}^{\star})\otimes \bm{h}^{\star}]
	\end{align}
 The first condition
	in (\ref{34}) is satisfied using the above definitions. Let us now define $\bm{\Phi}^{-1}=[\bar{\bm{L}} \bar{\bm{R}}]$, where $\bar{\bm{L}} \in \mathbb{C}^{5K\times K}$ and $\bar{\bm{R}} \in \mathbb{C}^{5k\times K}$. Then, we have 
\begin{align}
	\bar{\bm{\Gamma}}^{-1}=\bm{\bm{\Phi}}^{-1}\otimes \bm{I}_{L}=[\bar{\bm{L}} ~~\bar{\bm{R}}] \otimes \bm{I}_{L}= [\bar{\bm{L}}\otimes \bm{I}_{L} ~~ \bar{\bm{R}}\otimes \bm{I}_{L}],
\end{align}
and $\|\bar{\bm{\Gamma}}^{-1}\|=\|\bm{\bm{\Phi}}^{-1}\| \le 1.2301$. Then, we can use it to develop the following lemma.
	
	\begin{lem}\label{lem0}
On the event $\mathcal{E}_{1,\epsilon}$ with $\epsilon \in (0, 0.8189)$, we have
\begin{align}
	\|\bm{\Gamma}^{-1}-\bar{\bm{\Gamma}}^{-1}\|\le 2\|\bar{\bm{\Gamma}}^{-1}\|^{2}\epsilon, ~~ \|\bm{\Gamma}^{-1}\|\le \|\bar{\bm{\Gamma}}^{-1}\|.
\end{align}
The proof is the same with \cite[Corollary IV.5]{tang2013compressed} as it contains elementary linear algebra.
	\end{lem}
	
	\subsection{ Certifying (18b)}
	We show that $\bm{Q}(\bm{\tau})$ is concentrated around $\bar{\bm{Q}}(\bm{\tau})$ and then prove that $\|\bm{Q}(\bm{\tau})\|_{2}<1$ for $\bm{\tau} \notin \mathcal{T}$. Let us first define 
	\begin{align}
		&\bm{E}^{(\bm{i})}(\bm{\tau})= \kappa^{i_{1}+i_{2}+i_{3}+i_{4}}\nonumber\\& \bigg[\bm{K}^{(i_{1}, i_{2}, i_{3}, i_{4})}(\bm{\tau}-\bm{\tau}_{1}), \cdots, \bm{K}^{(i_{1}, i_{2}, i_{3}, i_{4})}(\bm{\tau}- \bm{\tau}_{K}), \nonumber\\
		&\bm{K}^{(i_{1}+1, i_{2}, i_{3}, i_{4})}(\bm{\tau}-\bm{\tau}_{1}), \cdots, \bm{K}^{(i_{1}+1, i_{2}, i_{3}, i_{4})}(\bm{\tau}- \bm{\tau}_{K}),\nonumber\\
		&\bm{K}^{(i_{1}, i_{2}+1, i_{3}, i_{4})}(\bm{\tau}-\bm{\tau}_{1}), \cdots, \bm{K}^{(i_{1}, i_{2}+1, i_{3}, i_{4})}(\bm{\tau}- \bm{\tau}_{K}),\nonumber\\
		&		\bm{K}^{(i_{1}, i_{2}, i_{3}+1, i_{4})}(\bm{\tau}-\bm{\tau}_{1}), \cdots, \bm{K}^{(i_{1}, i_{2}, i_{3}+1, i_{4})}(\bm{\tau}- \bm{\tau}_{K}),\nonumber\\
		&\bm{K}^{(i_{1}, i_{2}, i_{3}, i_{4}+1)}(\bm{\tau}-\bm{\tau}_{1}), \cdots, \bm{K}^{(i_{1}, i_{2}, i_{3}, i_{4}+1)}(\bm{\tau}- \bm{\tau}_{K})
		\bigg]^{T},
	\end{align}
then, we can write 
\begin{align}
	\bm{E}^{(\bm{i})}(\bm{\tau})&= \frac{1}{L^{4}}\sum_{\bm{n}\in \mathcal{J}}(j2\pi\kappa)^{i_{1}+i_{2}+i_{3}+i_{4}}n_{1}^{i_{1}}n_{2}^{i_{2}}n_{3}^{i_{3}}n_{4}^{i_{4}}\nonumber\\
	&s_{n_{1}}s_{n_{2}}s_{n_{3}}s_{n_{4}}e^{j2\pi \bm{\tau}^{T}\bm{n}}\bm{v}_{\bm{n}}\otimes (\bm{d}_{\bm{n}}\bm{d}_{\bm{n}}^{H}) .
\end{align}
Consequently, the exception can be given by 
\begin{align}
	&\bar{\bm{E}}^{(\bm{i})} (\bm{\tau})= \mathbb{E} \{ \bm{E}^{(\bm{i})} (\bm{\tau})\}\nonumber\\&=\frac{1}{L^{4}}\sum_{\bm{n} \in \mathcal{J}} (j2\pi 
	\kappa)^{i_{1}+i_{2}+i_{3}+i_{4}}n_{1}^{i_{1}}n_{2}^{i_{2}}n_{3}^{i_{3}}n_{4}^{i_{4}}\nonumber\\
	&s_{n_{1}}s_{n_{2}}s_{n_{3}}s_{n_{4}}e^{j2\pi 
	\bm{\tau}^{T}\bm{n}} \bm{v}_{\bm{n}}\otimes \bm{I}_{L}= \bm{E}^{(\bm{i})}(\bm{\tau})\otimes \bm{I}_{L}.
\end{align}
By defining $\bm{\Gamma}^{-1}= [\bm{L}~\bm{R}]$ and $\bm{\Phi}^{-1}=[\bar{\bm{L}}~\bar{\bm{R}}]$, such that $\bm{R} \in \mathbb{C}^{5KL\times 4KL}$ and $\bm{L}\in \mathbb{C}^{5KL\times KL}$, we can write 
\begin{align}
&\kappa^{i_{1}+i_{2}+i_{3}+i_{4}} \bm{Q}^{(\bm{i})}(\bm{\tau})=[\bm{E}^{(\bm{i})}(\bm{\tau})]^{H}\bm{L}\bm{h}\nonumber\\&=[\bm{E}^{(\bm{i})}(\bm{\tau})-\bar{\bm{E}}^{(\bm{i})}(\bm{\tau})+\bar{\bm{E}}^{(\bm{i})}(\bm{\tau})]^{H}(\bm{L}-\bar{\bm{L}}\otimes \bm{I}_{L}+ \bar{\bm{L}}\otimes \bm{I}_{L})\bm{h}\nonumber\\
&=[\bar{\bm{E}}^{(i)}(\bm{\tau})]^{H}(\bar{\bm{L}}\otimes \bm{I}_{L})\bm{h}+\bm{I}_{1}^{(i)}+\bm{I}_{2}^{(i)}(\bm{\tau}),
\end{align}
where $\bm{I}_{1}^{(\bm{i})}(\bm{\tau})=[\bm{E}^{(\bm{i})}(\bm{\tau})-\bar{\bm{E}}^{(\bm{i})(\bm{\tau})}]^{H}\bm{L}\bm{h}$ and $\bm{I}_{2}^{(\bm{i})}=[\bar{\bm{E}}^{(\bm{i})}(\bm{\tau})]^{H}(\bm{L}-\bar{\bm{L}}\otimes \bm{I}_{L})\bm{h}$. Regarding $\bar{\bm{Q}}^{(\bm{i})}(\bm{\tau})=[\bar{\bm{E}}^{(\bm{i})}(\bm{\tau})]^{H}(\bar{\bm{L}}\otimes \bm{I}_{L})\bm{h}$, one can write 
\begin{align}
	\kappa^{i_{1}+i_{2}+i_{3}+i_{4}} \bm{Q}^{(\bm{i})}(\bm{\tau})&=	\kappa^{i_{1}+i_{2}+i_{3}+i_{4}} \bar{\bm{Q}}^{(\bm{i})}(\bm{\tau})\nonumber\\&+\bm{I}_{1}^{(i)}(\bm{\tau})+\bm{I}_{2}^{(i)}(\bm{\tau}).
\end{align}
We now show that  $\|\bm{I}_{1}^{(i)}(\bm{\tau})\|_{2}$ and $\|\bm{I}_{2}^{(i)}(\bm{\tau})\|_{2}$ are small for a point on on the grid set $\Omega_{grid}$, which can be used to show that $\kappa^{i_{1}+i_{2}+i_{3}+i_{4}}\bm{Q}(\bm{\tau})$ is concentrated on $\kappa^{i_{1}+i_{2}+i_{3}+i_{4}}\bar{\bm{Q}}(\bm{\tau})$ over $\Omega_{grid}$. Finally, this is extended for the whole interval  $[0,1)^{4}$.

To specify the value of $\|\bm{I}_{1}^{(i)}(\bm{\tau})\|_{2}$, in the following Lemma, which proof is given in Appendix \ref{app2}, we use the Bernstein's inequality to approximate the values of $\|\bm{E}^{(\bm{i})}(\bm{\tau})-\bar{\bm{E}}^{(\bm{i})}(\bm{\tau})\|$, then, bound $\|\big(\bm{E}^{(\bm{i})}(\bm{\tau})-\bar{\bm{E}}^{(\bm{i})}(\bm{\tau})\big)^{H}\bm{L}\|$. 

\begin{lem}\label{lem4}
Let us define $0<\epsilon<1$ and $\bm{\tau} \in [0, 1)^{4}$, if 
\begin{align}
	L^{4} \geq \frac{134000\times4^{2(i_{1}+i_{2}+i_{3}+i_{4})}\mu KT}{3\epsilon^{2}}\log\bigg(\frac{4KT+T}{\delta}\bigg),
\end{align}
with the probability at least $1-256\delta_{2}$, we have $\|E^{(i)}(\bm{\tau})-\bar{E}^{(i)}(\bm{\tau})\|\le \epsilon_{2}$ for $i_{1},i_{2},i_{3}= \{0, \cdots, 4\}$.
\end{lem}

\begin{lem}\label{lem5}
Let us assume that $\epsilon \in (0, 0.9]$ and $\bm{\tau} \in \Omega_{grid}$, if 
\begin{align}
	L^{4} \geq \frac{134000\times4^{2(i_{1}+i_{2}+i_{3}+i_{4})}\mu KT}{3\epsilon^{2}}\log\bigg(\frac{4KT+T}{\delta}\bigg),
\end{align}
then, 
\begin{align}
	\mathbb{P}\Bigg\{\sup_{\bm{\tau}\in \Omega_{\mathrm{grid}}}&\bigg\|\big(\bm{E}^{\bm{i}}(\bm{\tau})-\bar{\bm{E}}^{(i)}(\bm{\tau})\big)^{H}\bm{L}\bigg\|\geq 4 \epsilon , \nonumber\\
&\bm{i}= \{0, \cdots, 4\}^{4}\Bigg\}
\le \mathbb{P}(\epsilon^{c}_{1, \epsilon})+ 256 \delta_{2} |\Omega_{\mathrm{grid}}|.
\end{align}
\end{lem}
Under the above event, we can write
\begin{align}
	\|\bm{E}^{(\bm{i})}(\bm{\tau})-\bar{\bm{E}}^{(i)}(\bm{\tau})\|\le \sqrt{L}\|\bm{E}^{(\bm{i})}(\bm{\tau})-\bar{\bm{E}}^{(i)}(\bm{\tau})\|\le \sqrt{L}\epsilon.
\end{align}
Consequently, in the following lemma, we prove that $\|\bm{I}_{1}^{\bm{i}}(\bm{\tau})\|$ can be well controlled over the sent $\Omega_{\mathrm{grid}}$.
	
	\begin{lem}
		\label{lem6}
		Let us consider $0 < \delta < 1$ and a finite set $\Omega_{\mathrm{grid}}$ over $[0,1)^{4}$, if 
		\begin{align}
			L^{4} \geq \frac{192 \mu KT}{\epsilon}\log\bigg(\frac{10KT}{\delta}\bigg),
		\end{align}
		where $C$ is constant, then, we have 
		\begin{align}
			\mathbb{P}\Bigg\{\sup_{\bm{\tau}\in \Omega_{\mathrm{grid
			}}}\|\bm{I}_{1}^{\bm{i}}(\bm{\tau})\|_{2}\le \epsilon, \bm{i} =\{0, \cdots, 3\}^{4}\Bigg\}\geq 1- 256\delta.
		\end{align}
		\end{lem}
Proof of Lemma \ref{lem6} is given in Appendix \ref{appE}.	We then bound $\bm{I}_{2}(\bm{\tau})$ in the following. First, let us bound $\|(\bm{L}-\bar{\bm{L}}\otimes\bm{I}_{L})^{H}\bar{\bm{E}}^{(\bm{i})}(\bm{\tau})\|_{F}^{2}$ in the following lemma.

\begin{lem}\label{lem9}
As long as 
\begin{align}
	L^{4} \geq \frac{192 \mu KT}{\epsilon}\log\bigg(\frac{10KT}{\delta}\bigg),
\end{align}
where $C$ is constant, one can bound $\|\bm{I}_{2}^{\bm{i}}(\bm\tau)\|_{2}$ as 
\begin{align}
	\mathbb{P} \Big\{\|\bm{I}_{2}^{\bm{i}}(\bm\tau)\|_{2} \le \epsilon, ~~\bm{i} \in \{0, \cdots, 3\}^{4}\Big\} \geq 1-256\delta,
\end{align}
where the proof is given by Appendix \ref{appendixf}.
\end{lem}
Here, we extend the above derivations for the whole interval $[0, 1)^{4}$ by choosing the grid size properly. Let us define the following event as
\begin{align}
	\mathcal{E}_{3}=\bigg\{\|\kappa^{i_{1}+i_{2}+i_{3}+i_{4}}Q^{(\bm{i})}(\bm{\tau})-\kappa^{i_{1}+i_{2}+i_{3}+i_{4}}\bar{Q}^{(\bm{i})}(\bm{\tau})\bm{h}^{\star}\|\bigg\} \le \frac{\epsilon_{3}}{3},
\end{align} 
for $\bm{i} \in \{0, \cdots, 3\}^{4}$.

 Using the above lemmas, we can write the following lemma. 

\begin{lem}\label{lem10}
If 
\begin{align}
		L^{4} \geq C\mu KT \log\bigg(\frac{10KT}{\delta}\bigg),
\end{align}
the event $\mathcal{E}_{3}$ holds with the probability at least $1-\delta$ where $C$ is constant. The proof is based on  \cite[Lemma 8]{chi2016guaranteed}.
\end{lem}
Then, it can be proved that as long as the minimum separation holds and 
\begin{align}
	L^{4} \geq C\mu KT \log\bigg(\frac{10KT}{\delta}\bigg),
\end{align}
we can bound
\begin{align}
	&\|Q(\bm{\tau})\|_{2} < 1-C_{a}, ~~~\forall \tau \in T_{\mathrm{far}}, \nonumber\\ 
	&\|Q(\bm{\tau})\|_{2} \le 1-C_{b}M^{2}(\bm{\tau}-\bm{\tau}_{i})^{2}, ~~~ \forall T_{i},
\end{align}
where $C_{a}$ and $C_{b}$ are constant such that $\rho_{c}^{2}C_{b}\le C_{a}$, holds with probability at least $1-\delta$. The proof can be found in \cite[Lemma 8]{chi2016guaranteed}.

\appendices

\section{Proof of Proposition  \ref{firstpro}}
\label{app}
We first show that any $\bm{q}$ that satisfies (\ref{18})  is dual feasible. To do so, we can write 
\begin{align}
	\|\bm{U}\|_{\mathcal{A}}&\geq \|\bm{U}\|_{\mathcal{A}}\|\mathcal{X}^{\star}(\bm{q})\|_{\mathcal{A}}^{\star} \geq \langle \mathcal{X}^{\star}(\bm{q}), \bm{U} \rangle_{\mathbb{R}}\nonumber\\
	&=\sum_{k=1}^{K}\mathrm{Re}\{\alpha_{k}\mathrm{trace}(\bm{c}(\bm{\tau}_{k})\bm{h}_{k}^{H}\mathcal{X}^{\star}(\bm{q}))\}\nonumber\\
	&=\sum_{k=1}^{K}\mathrm{Re}\{\alpha_{k}^{\star}\bm{h}_{k}^{H}\bm{Q}(\bm{\tau}_{k})\}\nonumber\\&=\sum_{k=1}^{K}\mathrm{Re}\{\alpha_{k}^{\star}\mathrm{sign}(\alpha_{k})\}= \sum_{k=1}^{K}|\alpha_{k}| \geq \|\bm{U}\|_{\mathcal{A}},
\end{align}
which leads to $\langle \mathcal{X}^{\star}(\bm{q}), \bm{U} \rangle_{\mathbb{R}}=\|\bm{U}\|_{\mathcal{A}}$. Therefore, $\bm{q}$ and $\bm{U}$ are the dual and primary optimum solutions, respectively.

To show that $\bm{U}$ is the unique solution of the problem, let us assume that $\tilde{\bm{U}}=\sum_{k}\tilde{\alpha}_{k}\bm{c}(\tilde{\bm{\tau}}_{k})\tilde{\bm{h}}^{T}_{k}$ as another solution to the problem has contains some elements outside of the support $T$ (otherwise the solutions are coincided). Then, we can write 
\begin{align}
	&\langle \mathcal{X}^{\star}(\bm{q}), \tilde{\bm{U}} \rangle_{\mathbb{R}} \nonumber\\
	&\sum_{\tilde{\tau}_{k} \in T}\mathrm{Re}\{\tilde{\alpha}_{k}^{\star}\tilde{\bm{h}}_{k}^{H}\bm{Q}(\tilde{\bm{\tau}}_{k})\}+\sum_{\tilde{\tau}_{k} \notin T}\mathrm{Re}\{\tilde{\alpha}_{k}^{\star}\tilde{\bm{h}}_{k}^{H}\bm{Q}(\tilde{\bm{\tau}}_{k})\}\nonumber\\
	&\le \sum_{\tilde{\tau}_{k} \in T}\tilde{\alpha}_{k}\|\bm{Q}(\tilde{\bm{\tau}}_{k})\|_{2}\|\tilde{\bm{h}}_{k}\|_{2}+ \sum_{\tilde{\tau}_{k} \notin T}\tilde{\alpha}_{k}\|\bm{Q}(\tilde{\bm{\tau}}_{k})\|_{2}\|\tilde{\bm{h}}_{k}\|_{2}\nonumber\\
	& <\sum_{\tilde{\tau}_{k} \in T}\tilde{\alpha}_{k}+ \sum_{\tilde{\tau}_{k} \notin T}\tilde{\alpha}_{k}= \|\tilde{\bm{U}}\|_{\mathcal{A}},
\end{align}
which contradicts strong duality. Therefore, the optimal solution of problem is unique. Then, we can form a linear system similar to \cite[Appendix A]{yang2016super} to show that linearly independent conditions
are satisfied.

\section{Proof of Lemma \ref{lem2}}
\label{appendix1}
To prove  Lemma \ref{lem2}, let us first define $B$ as below
\begin{align}\label{65}
	&B:=\|\bm{S}_{\bm{n}}\|\nonumber\\&=\frac{1}{L^{4}}\|s_{n_{1}}s_{n_{2}}s_{n_{3}}s_{n_{4}}(\bm{v}_{\bm{n}}\bm{v}_{\bm{n}}^{H})\otimes (\bm{d}_{\bm{n}}\bm{d}_{\bm{n}}^{H}-\bm{I}_{L})\|\nonumber\\
	&\le \frac{1}{L^{4}} \max_{n_{1}, n_{2}, n_{3}, n_{4}} |s_{n_{1}}||s_{n_{2}}||s_{n_{3}}||s_{n_{4}}|\|\bm{v}_{\bm{n}}\|_{2}^{2} \max\{\|\bm{b}_{\bm{n}}\|_{2}^{2}, \|\bm{I}_{L}\|\}\nonumber\\
& \le \frac{1}{L^{4}} (K+4K(2\pi n\kappa)^{2}) \max\{\mu T, 1\}\le \frac{53 K \mu T}{L^{4}},
\end{align}
where the last inequality comes from the fact that $\max_{|n|\le 2M }(2\pi n\kappa)^{2} \le 13$ for $M \geq 4$ \cite{candes2014towards}.

Second, let us define $\sigma^{2}$ as 
\begin{align}
	\sigma^{2}&=\bigg\|\sum_{\bm{n}\in \mathcal{J}}\mathbb{E}\{\bm{S}_{\bm{n}}\bm{S}_{\bm{n}}\}\bigg\|\nonumber\\
	&=\frac{1}{L^{8}}\bigg\|\sum_{\bm{n}\in \mathcal{J}}\mathbb{E}\Big\{s_{n_{1}}^{2}s_{n_{2}}^{2}s_{n_{3}}^{2}s_{n_{4}}^{2}\Big[(\bm{v}_{\bm{n}}\bm{v}_{\bm{n}}^{H})\otimes(\bm{d}_{\bm{n}}\bm{d}_{\bm{n}}^{H}-\bm{I}_{L})\Big]\nonumber\\
	&\Big[(\bm{v}_{\bm{n}}\bm{v}_{\bm{n}}^{H})\otimes(\bm{d}_{\bm{n}}\bm{d}_{\bm{n}}^{H}-\bm{I}_{L})\Big]^{H}\Big\}\bigg\|\nonumber\\
	&=\frac{1}{L^{8}}\bigg \|\sum_{\bm{n}\in \mathcal{J}}s_{n_{1}}^{2}s_{n_{2}}^{2}s_{n_{3}}^{2}s_{n_{4}}^{2} \|\bm{v}_{n}\|_{2}^{2}(\bm{v}_{n}\bm{v}_{n}^{H})\nonumber\\
	&\otimes \mathbb{E} \bigg\{(\bm{d}_{\bm{n}}\bm{d}_{\bm{n}}^{H}-\bm{I}_{L})(\bm{d}_{\bm{n}}\bm{d}_{\bm{n}}^{H}-\bm{I}_{L})^{H}\bigg\}\bigg \| \nonumber\\
&=\frac{1}{L^{8}}\bigg \|\sum_{\bm{n}\in \mathcal{J}}s_{n_{1}}^{2}s_{n_{2}}^{2}s_{n_{3}}^{2}s_{n_{4}}^{2} \|\bm{v}_{n}\|_{2}^{2}(\bm{v}_{n}\bm{v}_{n}^{H})\nonumber\\
&\otimes \mathbb{E} \bigg\{\|\bm{d}_{\bm{n}}\|_{2}(\bm{d}_{\bm{n}}\bm{b}_{\bm{n}}^{H}-\bm{I}_{L})\bigg\}\bigg \| \nonumber\\
	& \le \frac{53K\mu T}{L^{8}} \bigg\| \sum_{\bm{n}\in \mathcal{J}} s_{n_{1}}^{2}s_{n_{2}}^{2}s_{n_{3}}^{2}s_{n_{4}}^{2} (\bm{v}_{n}\bm{v}_{n}^{H} \otimes \bm{I}_{L}) \bigg\|  \nonumber\\
	&\le \frac{53K\mu T}{L^{4}} \max_{\bm{n} \in \mathcal{J}} s_{n_{1}}s_{n_{2}}s_{n_{3}}s_{n_{4}} \bigg\| \sum_{\bm{n}\in \mathcal{J}} s_{n_{1}}^{2}s_{n_{2}}^{2}s_{n_{3}}^{2}s_{n_{4}}^{2} (\bm{v}_{n}\bm{v}_{n}^{H} \otimes \bm{I}_{L}) \bigg\|  \nonumber\\
	&\le   \frac{53K\mu T}{L^{4}} \|\bar{\bm{\Gamma}}\|
	 \le 
	 \frac{96K\mu T}{L^{4}},
\end{align}
where we used (\ref{65}) and Lemma \ref{lem} for bounding $\|\bar{\bm{\Gamma}}\|=\|\bm{\Phi}\|\le 1.811$. Using Bernstein's non-commutative inequality
\begin{align}
	\mathbb{P}\{\|\bm{\Gamma}-\bar{\bm{\Gamma}}\| \geq t\} \le 2 d \exp\Bigg(\frac{-t^{2}/2}{\sigma^{2}+Bt/3}\Bigg),
\end{align}
where $d$ is the dimension of the matrices, we can bound 
\begin{align}
	\mathbb{P}\{\|\bm{\Gamma}-\bar{\bm{\Gamma}}\| \geq \epsilon\} \le  10KT \exp\Bigg(\frac{-\epsilon^{2}/2}{\frac{96K\mu T}{L^{4}}+\frac{17 K \mu T\epsilon}{L^{4}}}\Bigg).
\end{align}
To have $\mathbb{P}\{\|\bm{\Gamma}-\bar{\bm{\Gamma}}\| \geq \epsilon\}\le \delta$, 
\begin{align}
	L^{4} \geq \frac{192 \mu KT}{\epsilon}\log\bigg(\frac{10KT}{\delta}\bigg),
\end{align}
which concludes the proof.

\section{Proof of lemma \ref{lem4}}
\label{app2}

Let us first define 
\begin{align}
	\bm{Y}_{\bm{n}}^{(\bm{i})}&=\frac{1}{L^{4}} (j2\pi n \kappa)^{i_{1}+i_{2}+i_{3}+i_{4}}n_{1}^{i_{1}}n_{2}^{i_{2}}n_{3}^{i_{3}}n_{4}^{i_{4}}s_{n_{1}}s_{n_{2}}s_{n_{3}}s_{n_{3}}\nonumber\\&e^{j2\pi\bm{\tau}^{T}\bm{n}}\bm{v}_{\bm{n}}\otimes(\bm{b}_{\bm{n}}\bm{b}_{\bm{n}}^{H}-\bm{I}_{L}),~~\nonumber\\&\bm{E}^{(\bm{i})}(\bm{\tau})-\bar{\bm{E}}^{(\bm{i})}(\bm{\tau})=\sum_{\bm{n}\in \mathcal{J}}\bm{Y}_{\bm{n}}^{(\bm{i})},
\end{align}
Regarding the fact that $\max_{n}\|s_{n}\|_{\infty}\le 1$, $|2\pi n\kappa| \le 4$, and $\|\bm{v}_{\bm{n}}\|_{2}^{2}\le 65K$, we can bound $\|\bm{Y}_{\bm{n}}^{(\bm{i})}\|$ as below
\begin{align}
\|\bm{Y}_{\bm{n}}^{(\bm{i})}\| &\le  \frac{1}{L^{4}}(4)^{i_{1}+i_{2}+i_{3}+i_{4}}\|\bm{v}_{\bm{n}}\|_{2}\|\bm{b}_{\bm{n}}\bm{b}_{\bm{n}}^{H}-\bm{I}_{L}\| \nonumber\\&\le \frac{1}{L^{4}}(4)^{i_{1}+i_{2}+i_{3}+i_{4}}\sqrt{65K}\max\{\mu T, 1\}\nonumber\\&\le  \frac{1}{L^{4}}2^{4+2(i_{1}+i_{2}+i_{3}+i_{4})}\sqrt{K}\mu T :=R
\end{align}
Then, 
\begin{align}
	&\bigg\|\sum_{\bm{n}\in \mathcal{J}}\mathbb{E}\bigg\{(\bm{Y}_{\bm{n}}^{(\bm{i})})^{H}\bm{Y}_{\bm{n}}^{(\bm{i})}\bigg\}\bigg\|= \frac{1}{L^{8}}\bigg\|\sum_{\bm{n}\in \mathcal{J}}\mathbb{E}\bigg\{\prod_{f=1}^{4}(|s_{n_{f}}|^{2}\nonumber\\&|2\pi n_{f}\kappa|^{2i_{k}})(\bm{v}_{\bm{n}}^{H}\otimes (\bm{b}_{\bm{n}}\bm{b}_{\bm{n}}^{H}-\bm{I}_{L}))	(\bm{v}_{\bm{n}}\otimes (\bm{b}_{\bm{n}}\bm{b}_{\bm{n}}^{H}-\bm{I}_{L}))
	\bigg\}\bigg\|\nonumber\\
	&\le \frac{1}{L^{8}}4^{2(i_{1}+i_{2}+i_{3}+i_{4})}\bigg\|\sum_{\bm{n}\in \mathcal{J}}\|\bm{v}_{\bm{n}}\|_{2}^{2}\mathbb{E}\{(\bm{b}_{\bm{n}}\bm{b}_{\bm{n}}^{H}-\bm{I}_{L})^{2}\}\bigg\|\nonumber\\
	&\le\frac{1}{L^{8}}4^{2(i_{1}+i_{2}+i_{3}+i_{4})}\mu T(4L+1)^{4}(65K)\nonumber\\&\le\frac{16640\mu T}{L^{4}}4^{2(i_{1}+i_{2}+i_{3}+i_{4})}=\sigma^{2}.
\end{align}
Now, let us use the matrix Bernstein inequality to bound the following event 
\begin{align}
	\mathbb{P}\bigg\{\bigg\|\sum_{\bm{n}\in \mathcal{J}}\bm{Y}_{\bm{n}}^{(\bm{i})}\bigg\|\geq \epsilon\bigg\}\le (4KT+T)\exp\bigg(\frac{-3\epsilon^{2}}{8\sigma^{2}}\bigg).
\end{align}
To have the failure probability less than $\delta_{2}$, we can write 
\begin{align}
	\log\bigg((4KT+T)\exp\bigg(\frac{-3\epsilon^{2}}{8\sigma^{2}}\bigg)\bigg)\le \log(\delta_{2})
\end{align}
and consequently leads to 
\begin{align}
	L^{4} \geq \frac{134000\times4^{2(i_{1}+i_{2}+i_{3}+i_{4})}\mu KT}{3\epsilon^{2}}\log\bigg(\frac{4KT+T}{\delta_{2}}\bigg).
\end{align}
Finally, by exploiting the union bound for $\bm{i}\in \{0,1,2,3\}^{4}$, one can write 
\begin{align}
		\mathbb{P}\bigg\{\bigg\|\sum_{\bm{n}\in \mathcal{J}}\bm{Y}_{\bm{n}}^{(\bm{i})}\bigg\|\geq \epsilon, \bm{i} \in \{0, 1, 2, 3\}^{4}\bigg\}\le 256\delta_{2},
\end{align}
which requires $
	L^{4} \geq \frac{134000\times4^{2(i_{1}+i_{2}+i_{3}+i_{4})}\mu KT}{3\epsilon^{2}}\log\bigg(\frac{4KT+T}{\delta_{2}}\bigg)$.
	
	\section{Proof of lemma \ref{lem5}}
	\label{app3}
	In Lemma \ref{lem4}, we showed that $\|\bm{E}^{(\bm{i})}(\bm{\tau})-\bar{\bm{E}}^{(\bm{i})}(\bm{\tau})\| \le \epsilon$ with probability at least $1-256\delta$, required  $
		L^{4} \geq \frac{134000\times4^{2(i_{1}+i_{2}+i_{3}+i_{4})}\mu KT}{3\epsilon^{2}}\log\bigg(\frac{4KT+T}{\delta_{2}}\bigg)$. By considering $\mathcal{E}_{1, \epsilon}$ and $\epsilon \in [0, 0.9)$ and 
	\begin{align}
		\cap_{\bm{\tau} \in \Omega_{\mathrm{grid}}}\{\|\bm{E}^{(\bm{i})}(\bm{\tau})-\bar{\bm{E}}^{(\bm{i})}(\bm{\tau})\|\le \epsilon, \bm{i} \in \{0, 1, 2, 3\}^{4}\},
	\end{align}
we can write
\begin{align}
&\|\big(\bm{E}^{(\bm{i})}(\bm{\tau})-\bar{\bm{E}}^{(\bm{i})}(\bm{\tau})\big)^{H}\bm{L}\| \nonumber\\&\le \|\big(\bm{E}^{(\bm{i})}(\bm{\tau})-\bar{\bm{E}}^{(\bm{i})}(\bm{\tau})\big)^{H}\| \|\bm{L}\|\nonumber\\
&\le 2\epsilon\|\bar{\bm{\Gamma}}\|\le 4 \epsilon,
\end{align}
where the second inequality comes from the fact that $\bm{L}$ is a submatrix of $\bm{\Gamma}^{-1}$ and the last inequality is because of Lemma \ref{lem0}. Then, applying union bound leads to 
\begin{align}
	\mathbb{P}\Bigg\{\sup_{\bm{\tau}\in \Omega_{\mathrm{grid}}}&\bigg\|\big(\bm{E}^{\bm{i}}(\bm{\tau})-\bar{\bm{E}}^{(i)}(\bm{\tau})\big)^{H}\bm{L}\bigg\|\geq 4 \epsilon , \nonumber\\
	&\bm{i}= \{0, \cdots, 4\}^{4}\Bigg\}
	\le \mathbb{P}(\epsilon^{c}_{1, \epsilon})+ 256 \delta_{2} |\Omega_{\mathrm{grid}}|,
\end{align}
which concludes the proof.

\section{Proof of lemma \ref{lem6}}\label{appE}
	
	Let us define the event 
	\begin{align}
		\mathcal{E}_{2}&=\cap_{\bm{\tau} \in \Omega_{\mathrm{grid}}}\bigg\{\bigg\|\big(\bm{E}^{\bm{i}}(\bm{\tau})-\bar{\bm{E}}^{(i)}(\bm{\tau})\big)^{H}\bm{L}\bigg\|\nonumber\\
		&\le 2^{2(\bm{i})}T\sqrt{\frac{2\mu K}{L^{4}}}+a \bar{\sigma}_{\bm{i}}:=\lambda_{\bm{i}}	
		\bigg\}.
	\end{align}
Constrained on $\mathcal{E}_{1,\epsilon_{1}}$, we can write 
\begin{align}
	\|\bm{I}_{1}^{\bm{i}}(\bm{\tau})\|_{2}&\le \bigg\|\big(\bm{E}^{\bm{i}}(\bm{\tau})-\bar{\bm{E}}^{(i)}(\bm{\tau})\big)^{H}\bigg\|\|\bm{L}\|\|\mathrm{sign}(a)\otimes\bm{h}_{k}\|\nonumber\\&\le 4 \sqrt{K}\lambda_{\bm{i}},
\end{align}
where we used Lemma \ref{lem}. By using union bound  and setting $\lambda_{\bm{i}}\le \epsilon/(4\sqrt{K})$, one can write 
\begin{align}
	\mathbb{P}\bigg(\sup_{\bm{\tau}\in \Omega_{\mathrm{grid}}}\|\bm{I}_{1}^{\bm{i}}(\bm{\tau})\|\bigg)\le 256|\Omega_{\mathrm{grid}}|e^{-ca^{2}}+\mathbb{P}(\mathcal{E}_{1, \epsilon}^{c}),
\end{align}
where the second term $\mathbb{P}(\mathcal{E}_{1, \epsilon}^{c})\le \delta$ when $	L^{4} \geq \frac{192 \mu KT}{\epsilon_{1}}\log\bigg(\frac{10KT}{\delta}\bigg)$. By setting $a^{2}=c^{-1}\log(\frac{256|\Omega_{\mathrm{grid}}|}{\delta})$, leading to an upper bound for the first term as $\delta$, which concludes the proof.

\section{Proof of lemma \ref{lem9}}\label{appendixf}

To bound $\|\bm{I}_{2}^{\bm{i}}\|$, we know $\|\bar{\bm{E}}^{(\bm{i})}(\bm{\tau})\|_{F}\le C_{1}$ from []. Thus, $\|\bm{E}^{(\bm{i})}(\bm{\tau})\|\le C_{1}$. In addition, constrained on $\mathcal{E}_{1, \epsilon}$, we can write 
\begin{align}
	\|\bm{I}_{2}^{(\bm{i})}\|_{2} &\le \|\bm{E}^{(\bm{i})}(\bm{\tau})\|\|\bm{L}-\bar{\bm{L}}\otimes \bm{I}_{L}\|\|\mathrm{sign}(a^{\star})\otimes
	\bm{h}_{k}\|_{2}\nonumber\\&\le C^{\prime}\sqrt{K}\epsilon.
\end{align} 
To  bound this with $\epsilon_{1}$, set $\epsilon=\frac{\epsilon}{C^{\prime}\sqrt{K}}$. Then, using (\ref{mear}), we require $	L^{4} \geq \frac{192 \mu KT}{\epsilon_{1}}\log\bigg(\frac{10KT}{\delta}\bigg)$  for some large enough constant $C$.


	\bibliographystyle{IEEEtran}
	\bibliography{References}
\end{document}